\documentclass[prmaterials, reprint, aps, amsmath, amssymb, floatfix]{revtex4-2}
\usepackage{amsmath}
\usepackage{siunitx}
\usepackage{lmodern} 
\usepackage{mathtools}
\usepackage{changes}
\usepackage[bookmarks, colorlinks=false]{hyperref}
\usepackage{soul}
\hypersetup{colorlinks=true, allcolors=blue}

\newcommand{\revA}[1]{\textcolor{black}{#1}}
\renewcommand{\hl}[1]{#1}

\begin{document}

\title{Thermodynamics and dynamics of non-compact prismatic dislocation loops simulated using a machine-learning model}
\author{Sho Hayakawa}
	\email{sho.hayakawa@ukaea.uk}
	\affiliation{UK Atomic Energy Authority, Culham Campus, Abingdon, Oxfordshire OX14 3DB, United Kingdom}	
\author{Sergei L. Dudarev}
	\affiliation{UK Atomic Energy Authority, Culham Campus, Abingdon, Oxfordshire OX14 3DB, United Kingdom}
\author{Max Boleininger}
	\affiliation{UK Atomic Energy Authority, Culham Campus, Abingdon, Oxfordshire OX14 3DB, United Kingdom}	

\date{\today}
	
\begin{abstract}

We explore how the thermodynamic properties and dynamics of a self-interstitial prismatic dislocation loop are affected by microscopic-scale variations in its geometric configuration---an aspect that rarely received attention in literature. We develop a machine-learning (ML) model to predict the formation energy of an arbitrary geometrically complex configuration of a self-interstitial atom dislocation loop. Trained on atomistic simulation data, the ML model achieves high predictive accuracy across a broad range of configurations, with a typical error in the 1\% range. From the ML model, we evaluate the density of configurational microstates as a function of loop's formation energy and derive analytical expressions valid in tractable limiting cases. Using statistical mechanics, we derive the configurational free energy, the average energy, and the thermodynamic entropy of a dislocation loop as a function of temperature. We also simulate the dynamics of self-climb of dislocation loops with various geometries and evaluate their diffusion coefficients and effective activation energy for diffusion. Analysis shows that there is a single universal parameter describing the morphological irregularity of the loop configuration in its ground state. This parameter determines the thermodynamic properties of a loop as well as its dynamics, and simulations illustrate how the properties and mobility of a configurationally complex loop vary as functions of the irregularity parameter. 
   
\end{abstract}
\maketitle


\section{Introduction}
Structural materials in nuclear reactors are exposed to extreme environmental conditions, where irradiation by energetic particles generates a large number of nanoscale defects, such as self-interstitial atoms (SIAs) and vacancies \cite{Stoller2012}. The irradiation-induced defects migrate in the material which, through interaction between defects of similar or opposite type, leads to cluster formation or mutual annihilation. Simulations showed that defect clusters also form directly in displacement collision cascades \cite{Kirk1987JNuclMater,Malerba2006JNuclMater,Hayakawa2019JMaterSci54-11096}. Irradiation-induced defects are the features driving microstructural evolution, and responsible for the degradation of mechanical properties in reactor components, including irradiation embrittlement \cite{Singh1999JNuclMater,Zinkle2006JNuclMater} and dimensional instability \cite{Garner1984JNuclMater,Garner2000JNuclMater}. 

Some defect clusters form as two-dimensional, platelet-like structures known as prismatic dislocation loops. These loops can form directly in high-energy collision cascade events \cite{DelaRubia1991}, by the coalescence of point defects \cite{Chartier2019,Derlet2020} and, potentially, through a void to loop transformation \cite{Vijendran2024}. Whereas the fundamental properties of dislocation loops have been extensively studied, through the analysis of fundamental dislocation concepts  \cite{Kroupa,Teodosiu1982,indenbom_lothe_1992,Anderson2017} and by computational modelling \cite{Osetsky2000JNuclMater_276_65,Osetsky2000JNuclMater_276_202}, the particular aspects of loop behaviour and evolution, stemming from the strong irregularity of loop shapes and observed experimentally \cite{Mayoral2008,Wang2026},  have emerged recently from the studies of dislocation configurations formed under highly non-equilibrium conditions, see e.g. Refs. \cite{Sand2018,Boleininger2023}.  

For instance, the analytical models for the self-energy of loops, based on elasticity theory \cite{Wolfer2004PhysRevLett,Chu2011ActaMater}, assume smooth regular loop shapes. Atomistic modelling approaches, including density functional theory and empirical interatomic potentials, also assuming regular loop configurations, were employed to examine loop energetics, from which analytical expressions for the formation energy were derived and fitted to simulation results \cite{Ma2020PhysRevMaterials}. The dynamic behaviour of loops has likewise attracted considerable attention. Molecular dynamics studies suggested that certain types of SIA loops can undergo rapid one-dimensional migration along their Burgers vector, known as glide, and that this migration is characterised by very low activation energies which are almost independent of the loop size \cite{Osetsky2000JNuclMater_276_65}. Loops can also migrate three-dimensionally by moving perpendicular to their Burgers vector; 
however, this process is significantly slower than glide and involves a substantially higher activation energy \cite{Swinburne2016SciRep,Okita2016ActaMater}. Further research addressed the interaction of loops with other microstructural features, including other defect clusters \cite{Osetsky2000JNuclMater_276_202,Hayakawa2018PhilosMag} and dislocations \cite{Rodney2004ActaMater,Hayakawa2016NuclMaterEnergy9-581}. These studies provide a fundamental basis for understanding the properties and behaviour of dislocation loops. They also provide essential inputs to coarse-grained models for microstructural evolution under irradiation, including cluster dynamics \cite{Moll2013PhysRevLett,Jourdan2014JNuclMater}, object kinetic Monte Carlo \cite{Fu2005NatMater,Arevalo2007JNuclMater,Caturla2019_OKMC}, and mean-field approaches \cite{Messina2014PhysRevB}.

As we noted above, the existing models for dislocation loops primarily treat the relatively simple loop shapes, such as circles \cite{Alexander2016PhysRevB}, ellipses \cite{Wolfer2004PhysRevLett,Chu2011ActaMater}, perfect squares \cite{Dudarev2008PhysRevLett,Gao2022JNuclMater,Ma2020PhysRevMaterials} (not a typical dislocation loop shape that nevertheless is observed experimentally \cite{Yao2010PhilosMag,Inoue2026}), rectangles \cite{Alexander2016PhysRevB}, or hexagons \cite{Gao2022JNuclMater,Alexander2016PhysRevB,Ma2020PhysRevMaterials,Osetsky2000JNuclMater_276_202,Osetsky2000JNuclMater_276_65,Hayakawa2016NuclMaterEnergy9-581}. 
\revA{These approaches provide considerable flexibility in representing the dislocation loop geometries, and are highly efficient in the context of dislocation dynamics simulations \cite{Bako2011PhilosMag} as well as numerical solutions of  elasticity problems \cite{Gao2015JMPS,SantosGuemes2024JNuclMater}. However, the continuum elastic description and a smooth representation of dislocation lines might become problematic when atomic-scale variations in the line geometry are important \cite{Boleininger2020PhysRevResearch}.}
\revA{Furthermore, most of the existing} models assume a simply connected geometry where all the SIAs/vacancies composing a prismatic dislocation loop are continuously connected and there are no gaps or holes in the loop habit plane. Such simplifications are certainly advantageous, as they enable a straightforward representation of loops in coarse-grained models. However, in real microstructural environments, dislocation loops often exhibit far more complex morphologies, cf. the observed loop configurations in irradiated tungsten \cite{Ferroni2015,Wang2026}. Computational studies of displacement cascades also report the formation of irregularly shaped dislocation loops \cite{Hayakawa2019JMaterSci54-11096,Boleininger2022PhysRevMaterials,Voskoboinikov2008JNuclMater}. Subsequent simulations show that these irregular loops evolve dynamically, continuously changing their geometric configurations \cite{Hayakawa2023ComputMaterSci218-111987}. Experimental observations also reveal that gliding SIA loops can absorb vacancies at their perimeter \cite{Arakawa2020NatMater}. Such absorption processes drive the development of complex geometries and could even lead to the formation of disconnected loop segments. Moreover, the irregularity of loop geometry plays the central part in the self-climb process, where pipe diffusion is preferentially activated or enhanced at jogs along the loop perimeter \cite{Okita2016ActaMater,Swinburne2016SciRep}. These findings highlight the fact that accurate modelling of loop properties and behaviour requires an explicit consideration of the detailed geometry of the loops on the nanoscale, emphasizing a large gap between the simplified representations of loops adopted in existing models and the complex loop microstructure observed in real materials \cite{Ferroni2015,Wang2026}. Incorporating such complexity is essential for realistic simulations of microstructural evolution under irradiation, pivotal to the predictive capability and realism of multi-scale modelling frameworks.

In this study, we examine how atomic-scale variations in the dislocation loop geometry affect the thermodynamic properties and dynamics of a prismatic dislocation loop. As a first step, we develop predictive models for the formation energy of an arbitrary configuration of an SIA cluster, using machine learning parameterised to atomistic simulation data. We focus on a $\langle111\rangle$ self-interstitial prismatic dislocation loop in body-centred cubic (bcc) W, constituting the dominant-type SIA configuration in this material \cite{Dudarev2018PhysRevMaterials,Boleininger2022PhysRevMaterials}. The predictive accuracy of the models is validated across a broad range of configurations, yielding typical errors within 1\%. Using the model, we perform two types of simulations. First, we evaluate the density of geometric microstates for various loop configurations as a function of its formation energy. From the density of states, using equilibrium statistical mechanics, we evaluate the configurational free energy, the mean energy, and the entropy of loops as functions of temperature. Then, we simulate the loop dynamics involving self-climb, where the configurational complexity plays a central role. The diffusion coefficient and effective activation energy for self-climb are evaluated. Our key finding is that there is a universal parameter describing the irregularity of the loop geometry, and it is this parameter that determines both the thermodynamic properties of loops and their dynamics.

    \begin{figure*}[htbp]
    \centering
    \includegraphics[width=1.0\textwidth]{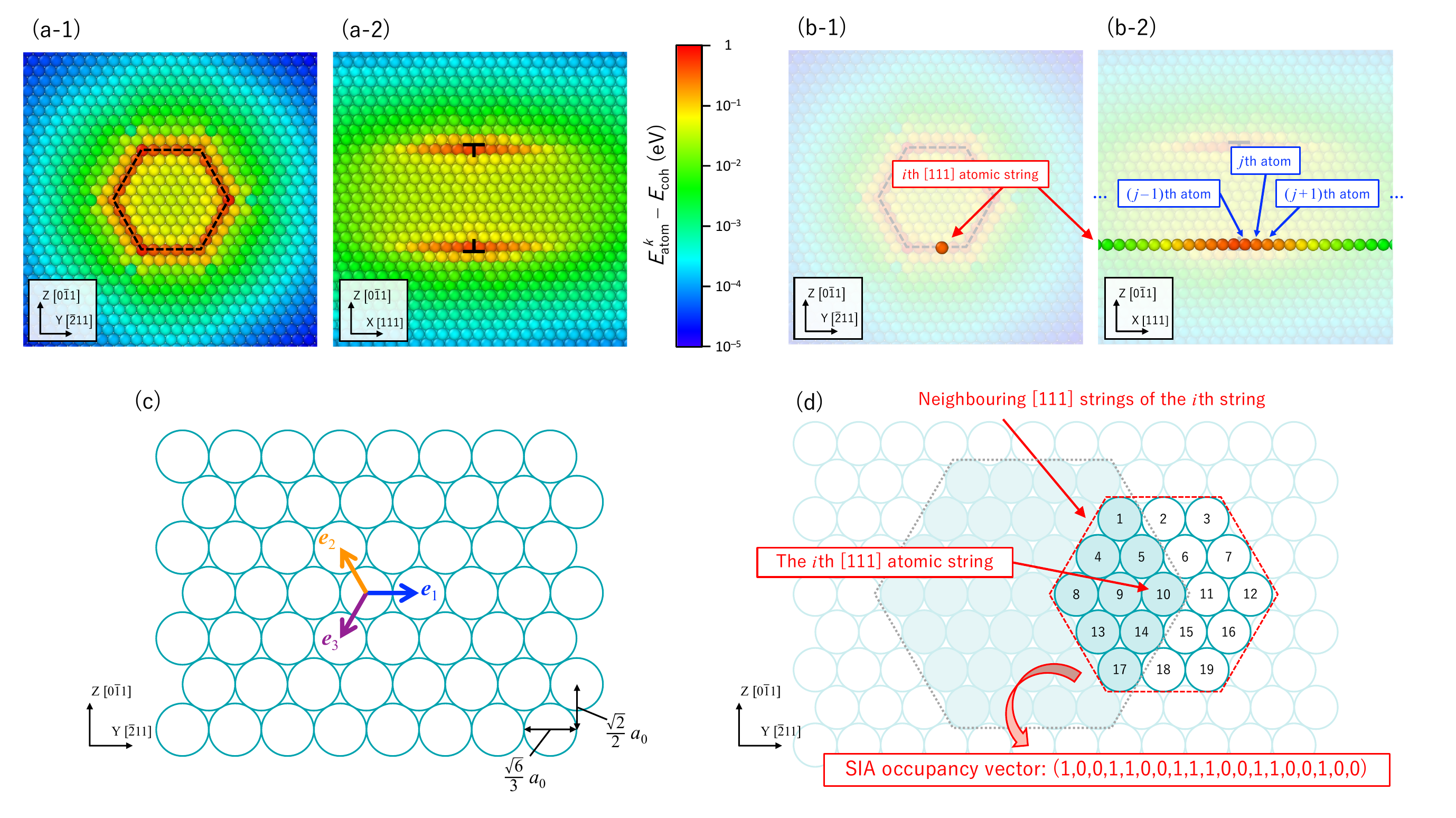}
    \caption{\revA{(a-1) and (a-2) Cross-sectional views of the simulation cell containing a hexagonal 91-SIA loop, taken in the $(111)$ and $(\bar{2}11)$ planes and passing through the loop centre, respectively. The atoms are coloured according to the energy increment due to the presence of the loop, $E^{k}_{\mathrm{atom}} - E_{\mathrm{coh}}$, as defined in Eq.~(\ref{eq:1}). The dashed black hexagon in (a-1) indicates the loop boundary, while the black symbols in (a-2) indicate the loop edges. (b-1) and (b-2) Cross-sectional views corresponding to (a-1) and (a-2), respectively, with a representative $i$th [111] atomic string highlighted. In (b-2), three consecutive atoms contained in the $i$th string---the $(j-1)$th, $j$th, and $(j+1)$th atoms---are indicated by blue arrows. (c) Two-dimensional coordinate system corresponding to a (111) plane. (d) Example of the atomic strings for a hexagonal 37-SIA loop viewed from the perspective of the [111] direction, illustrating the strings neighbouring a given $i$th string for cutoff $n_\mathrm{cut} = 2$ by a dashed red hexagon. Filled circles represent atomic strings containing an SIA, while open circles show strings without an SIA. The $i$th string and its neighbouring strings are labelled according to the corresponding component positions in the SIA-occupancy vector; for the configuration shown, the corresponding vector is (1,0,0,1,1,0,0,1,1,1,0,0,1,1,0,0,1,0,0). The grey dotted hexagon denotes the loop boundary.}}
    \label{fig:model_illustration}
    \end{figure*}


\section{Machine-learning model for a prismatic dislocation loop}

In this study, we investigate the geometric complexity of prismatic $|{\boldsymbol{b}}|=(a_0/2)\langle 111\rangle$ dislocation loops, which are two-dimensional planar arrangements of SIAs with the Burgers vector $\boldsymbol{b}$ normal to a \{111\}-type crystallographic plane in a bcc lattice. Here, $a_0$ denotes the lattice constant. These prismatic dislocation loops form under irradiation by the self-agglomeration of SIA defects \cite{Boleininger2022PhysRevMaterials}. The individual SIA defects forming these dislocation loop structures adopt $(a_0/2)\langle 111\rangle$ crowdion configurations elongated in a $\langle 111\rangle$ direction, with the strain field of individual crowdions becoming more delocalized at greater distances to the loop perimeter \cite{Dudarev2003,Boleininger2018}. In addition to fully connected dislocation loop configurations, the model is constructed to include configurations where crowdions are detached from the loop and dispersed in space in the same \{111\} plane. This approach enables the development of a general model capable of handling arbitrary configurations of a set of SIAs, regardless of whether an SIA set has a simply connected geometry or not, or if the loop contains missing crowdions or even large holes in its plane. 

\revA{Given the relative simplicity of planar two-dimensional dislocation configurations considered below, the model does not explicitly treat the atomic vibrational degrees of freedom. The energies of different configurations are separated by the relatively large energy gaps, and the hierarchy of energy scales resembles that of molecules \cite{LandauQM}. Each dislocation and defect configuration is characterised by its own density of phonon states, where the differences between the densities of phonon states of different configurations reflect variations in interatomic interactions at the most strongly distorted lattice sites \cite{Lifshitz1966,Lapointe2025}. The vibrational free energy of dislocation loops also includes contributions from the soft modes associated with the glide motion of loops \cite{Osetsky2000JNuclMater_276_65}. We do not treat these low energy excitations, and also do not include contributions from the spontaneous thermal formation of vacancies \cite{Kraftmakher1998}. Appreciating the limitations arising from these approximations, we focus on how to treat the configurational degrees of freedom of dislocation loops {\it in principle}. When complex dislocation structures form in highly non-equilibrium cascade events initiated by high-energy atomic impacts typical of irradiation environments, it is the relative statistical weights of various realisations of dislocation structures that dominate their occurrence. The analysis of statistical configurations of loops below therefore reflects the largest, configurational, energy scales of dislocation loop ensembles, not treated in literature so far, whereas the full analysis of the problem must also include the finer energy scales associated with vibrational, electronic and, where appropriate, magnetic degrees of freedom \cite{Kormann2008}.}      

\revA{A rigorous treatment of vibrational contributions is known to be particularly challenging for large and complex defect configurations. The vibrational free energy itself is configuration dependent and needs to be evaluated separately for each thermally relevant configuration. For systems with large configurational spaces, such a configuration-resolved treatment rapidly becomes computationally prohibitively expensive \cite{Tolborg2026CrystEngComm}. This challenge is reflected in the existing literature, where multi-million atomic configurations containing hundreds of defect sites are normally not addressed. Most first-principles studies of vibrational free energies of defects have focused on isolated point defects or relatively small defect complexes, for which configuration-resolved lattice-dynamics calculations remain tractable \cite{Reith2009PhysRevB,Kong2025AIPAdv,Lucas2009NIMB}. Similar difficulties arise in chemically disordered alloys, where numerous distinct local chemical environments generate extensive configurational space and strongly site-dependent defect free energies \cite{Wrobel2025PhysRevB,Xu2026ScrMater}. Some studies have addressed the vibrational properties of self-interstitial atom clusters; however, these studies used restricted configurational sampling or additional approximations, for example an assumption that configurations with similar potential energies had similar vibrational characteristics \cite{Kapur2010PhysRevB,Chuang2015JApplPhys}. These limitations highlight the current difficulty of systematically resolving the vibrational free energies over a very large number of thermally accessible configurations characterising complex defect clusters.}

\revA{Having clarified the physical scope and limitations of the present model, we now proceed to its explicit construction.}
If SIAs are placed in a perfect crystal, their formation energy ($E_{\mathrm{f}}$) is given by the sum of all the energy increments in the system due to the presence of SIA defects:
    \begin{equation}
    E_{\mathrm{f}} = \sum_k \left( E^{k}_{\mathrm{atom}} - E_{\mathrm{coh}} \right) \text{,} \label{eq:1}
    \end{equation}
where $E^{k}_{\mathrm{atom}}$ and $E_{\mathrm{coh}}$ denote the energy of the $k$th atom and the atomic cohesive energy, respectively, \revA{and the summation runs over all atoms in the system (see the example of a 91-SIA loop shown in Fig.~\ref{fig:model_illustration}~(a-1) and (a-2))}. Alternatively, by grouping the atoms into individual $[111]$ atomic strings \cite{Dudarev2003}, $E_{\mathrm{f}}$ can be rewritten as:
    \begin{equation}
    E_{\mathrm{f}} = \sum_i E^{i}_{[111]} \text{,} \label{eq:2}
    \end{equation}
with
    \begin{equation}
    E^{i}_{[111]} = \sum_{j \in i\,[111]} \left( E^{j}_{\mathrm{atom}} - E_{\mathrm{coh}} \right) \text{,} \label{eq:3}
    \end{equation}
where $i\,[111]$ is the set of atoms belonging to the $i$th [111] atomic string, $E^{i}_{[111]}$ represents the sum of the energy increments over the atoms in the $i\,[111]$ set, and $E^{j}_{\mathrm{atom}}$ is the energy of the $j$th atom, see \revA{Fig.~\ref{fig:model_illustration}~(b-1) and (b-2)}. In this way, the formation energy $E_{\mathrm{f}}$ is decomposed into contributions from individual [111] atomic strings \cite{Dudarev2003,Boleininger2018}. Each term $E^{i}_{[111]}$ depends on the local environment of the $i$th atomic string, i.e. on the pattern of SIA occupancies of neighbouring strings. Based on this concept, we develop a model that predicts $E^{i}_{[111]}$ from local SIA occupancy patterns using machine-learning, and subsequently obtain $E_{\mathrm{f}}$ by summing the contributions from all the  $[111]$ atomic strings in the system. To make the model tractable, we assume that all the SIAs are aligned in the same direction in the lattice, and that no atomic string contains more than one SIA defect at a time. 

    \begin{figure*}[t]
    \centering
    \includegraphics[width=1.0\textwidth]{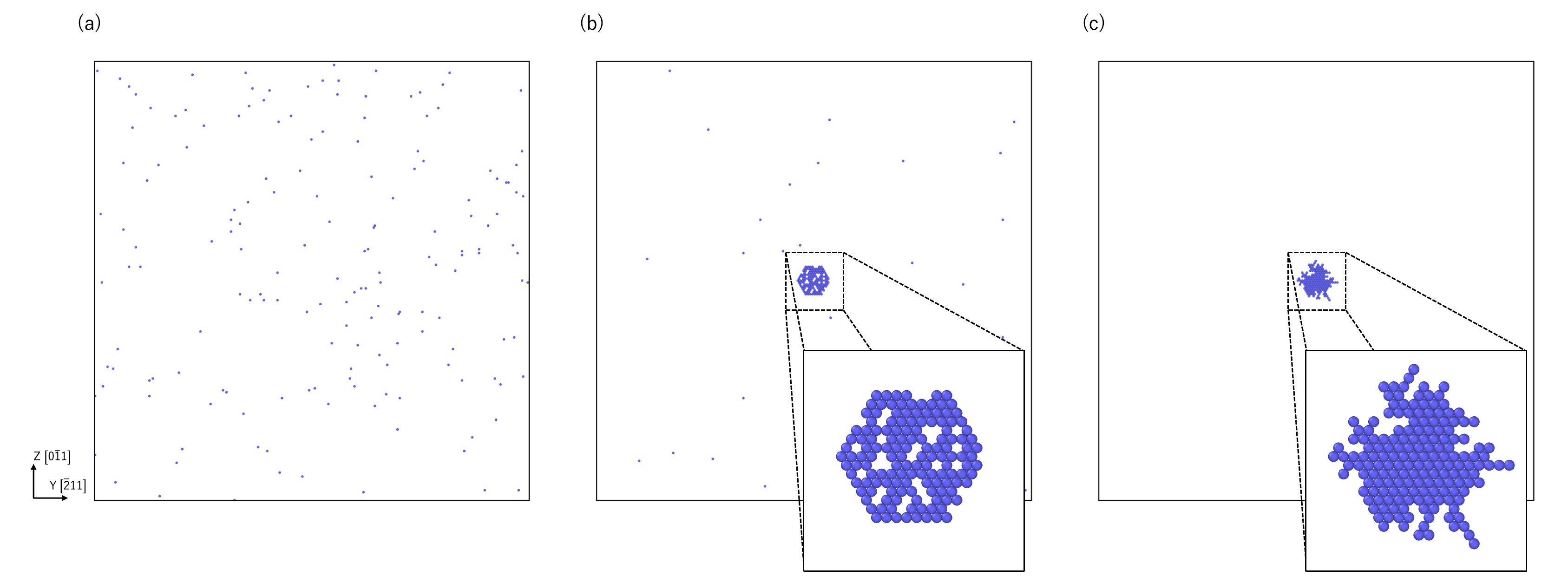}
    \caption{Examples of the generated SIA configurations ($N_{\mathrm{SIA}} = 169$). (a) Configuration generated by the first approach. (b) Configuration after displacing the 30th SIA by the second approach. (c) Configuration after displacing the 30th SIA by the third approach.}
    \label{fig:create_various_configurations}
    \end{figure*}
    
A local occupancy pattern is defined by the SIA occupancies of the $i$th and its neighbouring strings. To specify the neighbourhood range, we introduce a cutoff length $n_\mathrm{cut}$ as follows. We use a two-dimensional coordinate system in the (111) plane, defined by the following three basis vectors, see Fig.~\ref{fig:model_illustration}~(c):
    \begin{align}
    \boldsymbol{e}_1 &= \left( \frac{\sqrt{6}}{3} a_0 \right) \hat{\boldsymbol{y}} \text{,} \label{eq:e4} \\ 
    \boldsymbol{e}_2 &= -\frac{1}{2} \left( \frac{\sqrt{6}}{3} a_0 \right) \hat{\boldsymbol{y}} + \left( \frac{\sqrt{2}}{2} a_0 \right) \hat{\boldsymbol{z}} \text{,} \label{eq:e5} \\
    \boldsymbol{e}_3 &= -\frac{1}{2} \left( \frac{\sqrt{6}}{3} a_0 \right) \hat{\boldsymbol{y}} - \left( \frac{\sqrt{2}}{2} a_0 \right) \hat{\boldsymbol{z}} \text{.} \label{eq:e6}
    \end{align}
Here, $\hat{\boldsymbol{y}}$ and $\hat{\boldsymbol{z}}$ denote the unit vectors along the Y and Z axes, respectively. In this coordinate system, the position of the $i$th atomic string is represented by an integer vector $\boldsymbol{n}_i = (n_{i,1}, n_{i,2}, n_{i,3}) \in \mathbb{Z}^3$ as:
    \begin{equation}
    \boldsymbol{r}_i = \sum_k n_{i,k} \boldsymbol{e}_k \text{.}
    \end{equation}
The set of three integer numbers defining the position of an atomic string in the $(111)$ plane $\boldsymbol{n}_i$ is not uniquely defined due to the redundancy of the basis vectors. Indeed, the two-dimensional projections of [111] atomic strings onto the (111) plane form a two-dimensional hexagonal Bravais lattice, where the sites can be parametrised using a set of two non-collinear basis vectors similar to those used in surface crystallography \cite{VanHove1986}. Our convention has the advantage of computational efficiency and it also explicitly reflects the symmetry of the problem. To define $\boldsymbol{n}_i$ uniquely, we impose a constraint on the set of integers:
    \begin{equation}
    \boldsymbol{n}_i = \arg \min_{\boldsymbol{m}_i \in \mathbb{Z}^3 : \sum m_{i,k} \boldsymbol{e}_k = \boldsymbol{r}_i} \| \boldsymbol{m}_i \|_1 \text{,} \label{eq:8}
    \end{equation}
where $\| \boldsymbol{m}_i \|_1$ denotes the $L^1$ norm of $\boldsymbol{m}_i$. This means that, among all the possible integer representations of $\boldsymbol{n}_i$, we choose the one that minimises the $L^1$ norm of $\boldsymbol{n}_i$, see Sec.~I of Supplementary Materials for a proof of the uniqueness of $\boldsymbol{n}_i$, defined by Eq.~(\ref{eq:8}). The $j$th string is considered ``neighbouring'' to the $i$th string if the $L^1$ norm of $\boldsymbol{n}_j$ \textit{expressed with the origin shifted to the $i$th string position} ($\| \boldsymbol{n}_{(j|i)} \|_1$) does not exceed the value of $n_\mathrm{cut}$:
    \begin{equation}
    \| \boldsymbol{n}_{(j|i)} \|_1 \leq n_\mathrm{cut} \text{.}
    \end{equation}
For instance, the red dashed-line hexagon in Fig.~\ref{fig:model_illustration}~(d) represents a region encompassing the neighbouring atomic strings for the $i$th string when $n_\mathrm{cut} = 2$. 
    
\revA{A variety of approaches have been proposed to encode information about local environments as descriptors for machine-learning models. For instance, the smooth overlap of atomic positions (SOAP) descriptor represents the local atomic density through an expansion in radial basis functions and spherical harmonics \cite{Bartok2013PhysRevB}. Similarly, atom-centred symmetry functions (ACSFs) describe local atomic environments using a set of radial and angular functions \cite{Behler2011JChemPhys}. These descriptors have been widely used in machine-learning interatomic potentials \cite{Szlachta2014PhysRevB,Dragoni2018PhysRevMaterials,Artrith2012PhysRevB}. In contrast, direct discrete representations of local lattice environments have also been employed in machine-learning models. Xu \textit{et al.} encoded the atomic species occupying lattice sites around a point defect as binary digits, which are assembled into a binary vector and supplied directly to the input layer of a neural network \cite{Xu2022ActaMater,Xu2022MatDes,Xu2023CellRepPhysSci}. This approach has enabled accurate predictions of point-defect migration energy barriers, including in concentrated multi-principal element alloys, where the local energy landscape for point defect diffusion is highly heterogeneous \cite{Xu2023CellRepPhysSci}. Another example is provided by Kimari \textit{et al.}, who represented the atomic occupation states of surface lattice sites using binary digits and directly supplied them to a neural network for predicting surface-diffusion barriers \cite{Kimari2020CompMaterSci}. A notable feature of this class of approaches is that the required information about the local environment---namely, atomic species or occupation states---is inherently discrete. Consequently, the local environment can be represented using a simple binary encoding without requiring a continuous description of the atomic configuration. The present problem has a closely analogous structure. Here, the local environment of an atomic string is determined by the presence or absence of an SIA on neighbouring atomic strings. This information is likewise discrete and therefore does not require a continuous representation of the local environment. Furthermore, the target quantity in the present model is the energy contribution of an individual atomic string. Accordingly, this quantity can be inferred directly from the discrete information describing the near atomic environment. Based on this analogy, we encode the occupancy of each neighbouring atomic string using a binary value, with 1 and 0 indicating the presence and absence of an SIA in a given atomic string, respectively. These binary values are assembled into an SIA-occupancy vector and supplied directly to the input layer of the neural network. As an example, Fig.~\ref{fig:model_illustration}~(d) illustrates the construction of the SIA-occupancy vector for $n_\mathrm{cut} = 2$. The $i$th atomic string and its neighbouring strings are labelled 1--19 according to their corresponding component positions in the occupancy vector. For the configuration shown, the resulting occupancy vector is (1,0,0,1,1,0,0,1,1,1,0,0,1,1,0,0,1,0,0). As each vector component is associated with a specific relative position of an atomic string, different spatial arrangements are represented by different occupancy vectors. We note that the ordering of the vector components is an arbitrary indexing convention and has no physical significance, provided that the correspondence between each spatial position and its associated input node is defined consistently.}


    
The training and test datasets are constructed as follows. To generate various configurations of SIAs, we first prepare a perfect crystal cell oriented along the X [111], Y $[\overline{2}11]$, and Z $[0\overline{1}1]$ directions, with the cell dimensions of 50.4, 49.6, and {50.1\,nm}, respectively, \revA{containing a total of 7,913,472 atoms.} Periodic boundary conditions are applied in all directions. A set of $N_{\mathrm{SIA}}$ SIAs is then introduced in the YZ plane using three different occupancy selection approaches, for the total SIA numbers in the cell of $N_{\mathrm{SIA}} =$ 37, 91, and 169. In the first approach, the SIAs are randomly dispersed in the YZ plane, see Fig.~\ref{fig:create_various_configurations}~(a). A total of 500 distinct configurations are generated for each $N_{\mathrm{SIA}}$ by varying the random SIA positions. In the second approach, the SIAs are initially arranged in a hexagonal configuration. Then, one after another, SIAs are selected in a random order and displaced from their initial configuration to a random location within the same YZ plane, until all the SIAs are displaced from their initial positions, see Fig.~\ref{fig:create_various_configurations}~(b). Since a new configuration is obtained after each displacement, $N_{\mathrm{SIA}}$ distinct configurations are generated in addition to the initial hexagonal configuration. The entire process is repeated five times for each $N_{\mathrm{SIA}}$. In the third approach, the SIAs are initially arranged in a hexagonal configuration. One SIA is then randomly selected and displaced to a random position so that the geometry of the configuration maintains \textit{topological equivalence}, see Fig.~\ref{fig:create_various_configurations}~(c). Here, topological equivalence means that all the SIA-containing strings remain continuously connected and that no `holes' are introduced in the configuration. The displacement procedure is repeated 100 times, and the entire process is conducted five times for each $N_{\mathrm{SIA}}$.

For each generated configuration, all the possible local SIA-occupancy patterns are identified and stored in the dataset. 
\revA{Periodic boundary conditions are taken into account when identifying the neighbouring atomic strings, such that strings across a simulation-cell boundary are treated as neighbours through their periodic images.}
It should be noted that the central string position does not necessarily need to be an SIA-containing string. The local occupancy pattern for the $i$th atomic string is labelled with the corresponding energy according to Eq.~(\ref{eq:3}). The energy of each atom in the configurations is computed by minimising the total system energy using atomistic simulations. To achieve this, we employ a combination of static relaxation by the conjugate gradient method and quasi-dynamic relaxation with a low effective temperature of {10 K}, which allows the system to escape from shallow local traps on the potential energy surface. Calculations are performed using the large atomic/molecular massively parallel simulator (LAMMPS) \cite{Plimpton1995}, and the interatomic interactions are described by an embedded atom method-type potential for bcc W developed by Mason {\it et al.} \cite{Mason2017JPhysCondensMatter} Since the local SIA-occupancy patterns are characterised by the threefold rotational as well as inversion symmetries about the [111] axes centred at the $i$th position, the symmetrically equivalent patterns are generated for each sampled pattern and assigned the same energy. Duplicated patterns are excluded if they have already been included in the dataset. 

\revA{No pressure control is applied in the atomistic simulations for generating the training and test dataset, i.e., the cell dimensions are fixed throughout the simulations. This results in a residual stress in the simulation cell due to the introduction of SIA defects. For instance, for the cell size used here, the magnitude of the residual normal stress along the X $[111]$ direction reaches approximately 17 MPa for $N_{\mathrm{SIA}} = 169$. The residual stress arises because the fixed cell dimensions impose a zero macroscopic strain condition, thereby preventing the strain field generated by the SIA defects from being fully accommodated by an overall deformation of the simulation cell \cite{Ma2019PhysRevMaterials}. In addition, under periodic boundary conditions, each defect configuration interacts elastically with its periodic images. To assess the influence of these finite-size effects on $E_\mathrm{f}$, we have evaluated the elastic correction energy for each generated configuration, following Varvenne {\it et al.} \cite{Varvenne2013} and Ma \textit{et al.} \cite{Ma2020ComputPhysCommun}. This correction energy consists of the self-strain correction energy associated with the strain constraint of the fixed simulation cell and the elastic interaction energy between the defect and its periodic images. The resulting elastic correction energy was found to be at most 0.04\% of the total $E_\mathrm{f}$, confirming that the finite-size effects on $E_\mathrm{f}$ are negligible.}

    \begin{figure*}[t]
    \centering
    \includegraphics[width=0.95\textwidth]{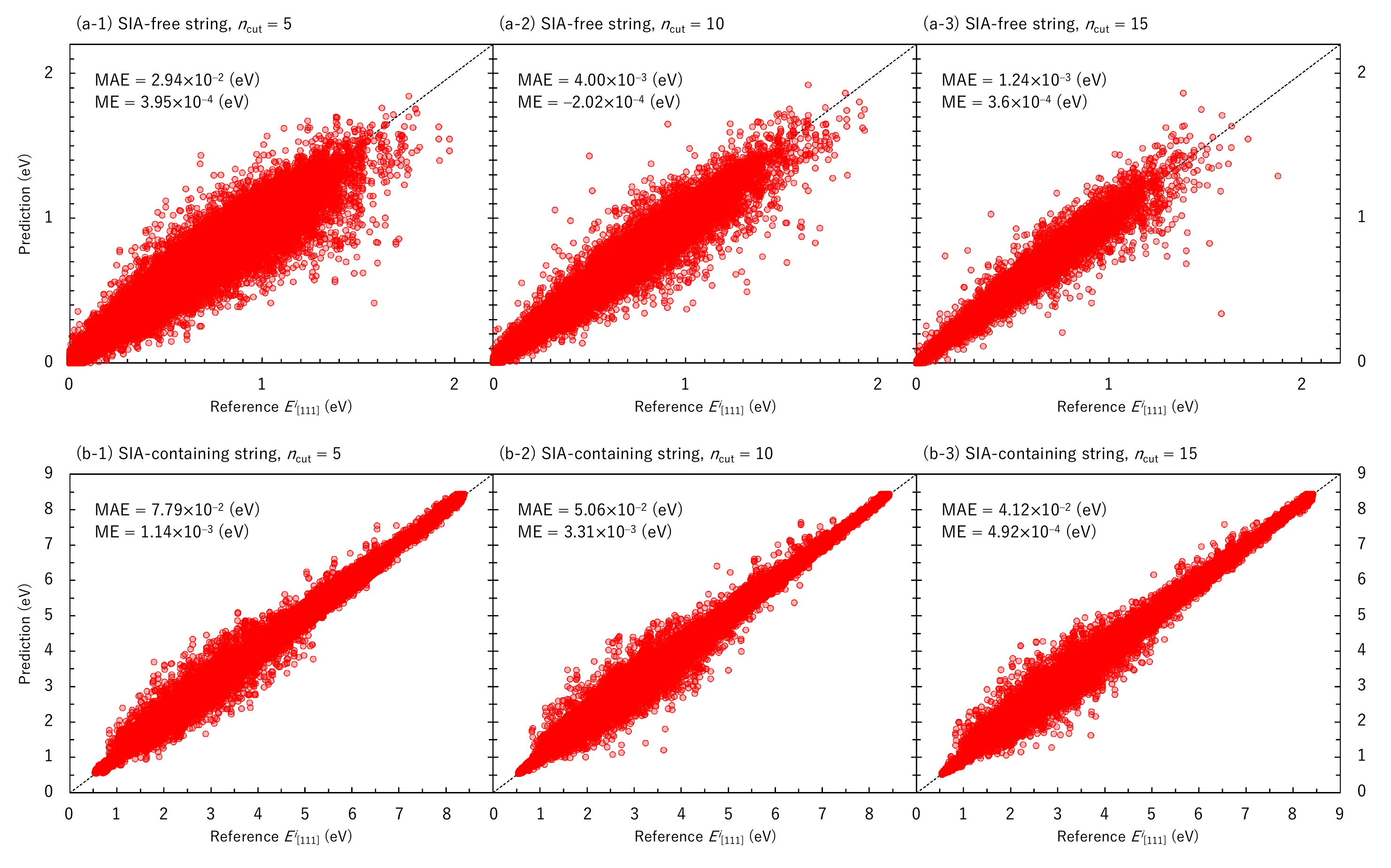}
    \caption{Values of $E^{i}_{[111]}$ predicted by the machine learning models \revA{using the test dataset} as functions of the corresponding reference values computed using atomistic energy minimisation. Panels (a-1)–(a-3) show results for SIA-free strings with $n_\mathrm{cut}$ = 5, 10, and 15, respectively. Panels (b-1)–(b-3) show the corresponding results for SIA-containing strings.}
    \label{fig:e111_prediction}
    \end{figure*}
    
We develop separate machine-learning models for the two cases---whether the $i$th atomic string contains an SIA or is SIA-free---because it is found that this yields better predictive capacity than the treatment of both cases within a single model. 
\revA{The value of $n_\mathrm{cut}$ is varied among 5, 10, and 15 to examine the sensitivity of the model performance to $n_\mathrm{cut}$. Accordingly, the dimensionalities of an SIA-occupancy vector are 91, 331, and 721 for $n_\mathrm{cut} =$ 5, 10, and 15, respectively.} As the number of local SIA-occupancy patterns sampled in each configuration depends on $n_\mathrm{cut}$, the number of generated data points also varies with $n_\mathrm{cut}$.
For the SIA-containing strings, the numbers of the generated data points are 669,234, 1,429,549, and 1,931,551 for $n_\mathrm{cut}=$ 5, 10, and 15, respectively, while in the SIA-free case, they are 1,313,527, 17,113,438, and 97,423,102, respectively. In each case, the data are randomly split into training (80\%) and test (20\%) sets. Feed-forward neural network models are constructed using TensorFlow \cite{Abadi2016tensorflow}. The models consist of an input layer followed by three fully-connected hidden layers with 256, 128, 64 units, respectively, each employing Rectified Linear Unit (ReLU) activation (see Sec.~II and Fig.~S1 in Supplementary Materials for a discussion of the architecture influence on model performance). For the output layer, while a single linear neuron is used for the SIA-free string models, we impose a physically-motivated upper bound on the output neuron for the SIA-containing string models. Here, the upper bound is set to the $E^{i}_{[111]}$ value of a completely isolated SIA-containing string surrounded by SIA-free strings, i.e., 8.421 eV. This constraint effectively prevents the models from producing unphysical, abnormally large values of $E^{i}_{[111]}$, which might arise due to limited extrapolation capability. Such non-physical outputs were also noted in previous studies using neural network interatomic potentials \cite{Lin2022ComputMaterSci,Tokita2023JChemPhys}. Our network is trained to predict values of $\ln ([E^{i}_{[111]}/{\rm eV}]+1)$, where $E^{i}_{[111]}/{\rm eV}$ is a numerical value of energy expressed in electron-volt units. The resulting logarithmic values are then converted back to the values of $E^{i}_{[111]}$. This approach allows the model to better accommodate a very broad range of target energy values. The training is performed using the Adam optimizer with a learning rate of $10^{-3}$, and the mean-square error is adopted as the loss function. Early stopping with a patience of 50 epochs is applied based on the validation loss to prevent over-fitting.
    

\section{Model validation}

    \begin{figure*}[t]
    \centering
    \includegraphics[width=0.95\textwidth]{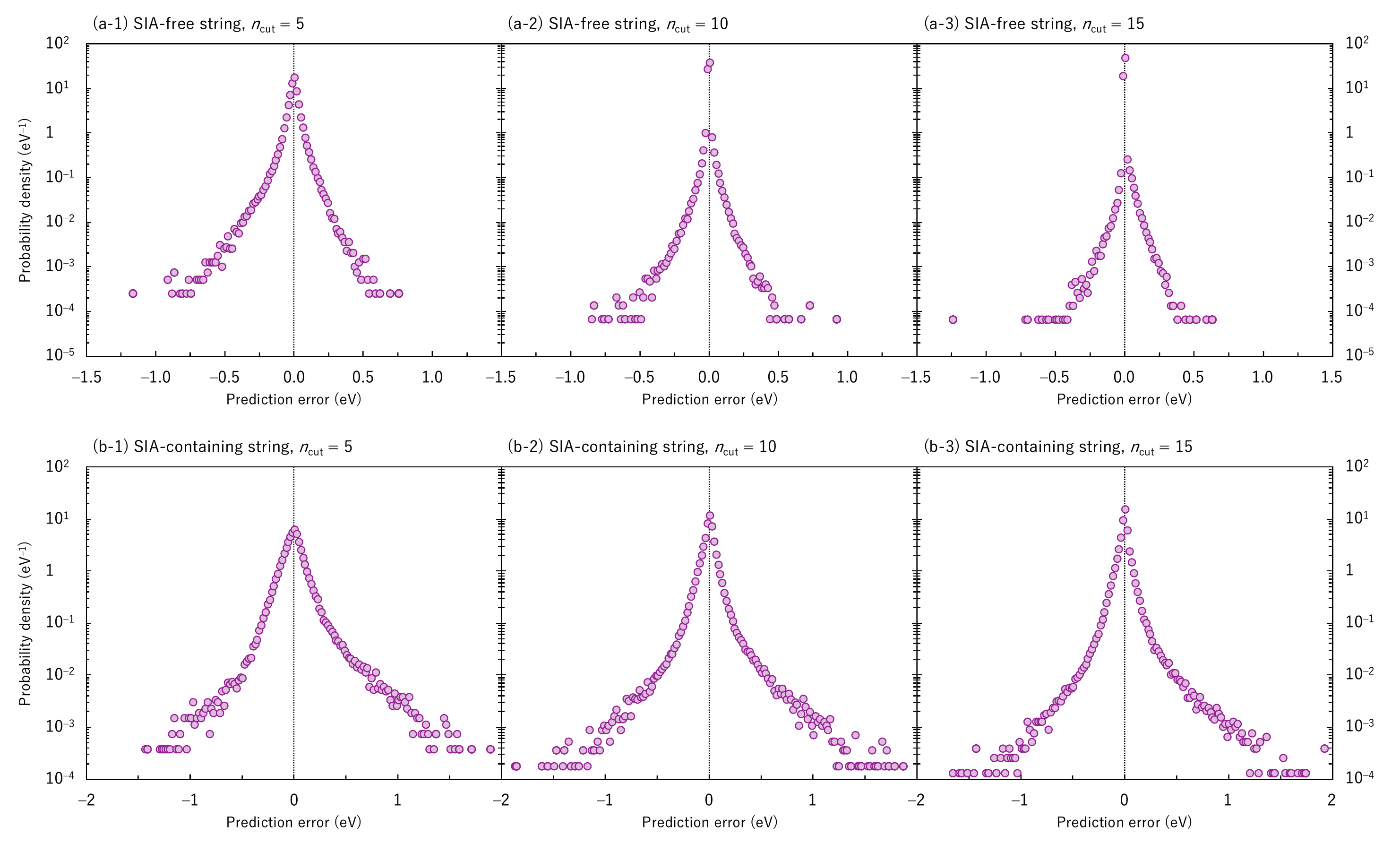}
    \caption{\revA{Probability density distributions of the prediction errors for the machine learning models, evaluated using the data points shown in Fig.~\ref{fig:e111_prediction}. Panels (a-1)--(a-3) show the results for SIA-free strings with $n_\mathrm{cut}$ = 5, 10, and 15, respectively. Panels (b-1)–(b-3) show the corresponding results for SIA-containing strings.}}
    \label{fig:probability_density}
    \end{figure*}
    
In this section, we assess the predictive accuracy of the machine-learning models developed above. Figs.~\ref{fig:e111_prediction}~(a-1)--(a-3) show the values of $E^{i}_{[111]}$ \revA{predicted using the test dataset} over the reference energy values obtained by energy minimisation of SIA-free strings for various $n_\mathrm{cut}$. From these results, the mean absolute error (MAE) and mean error (ME) of predictions are calculated and shown in each panel. 
\revA{The magnitude of the ME is substantially smaller than the corresponding MAE and remains close to zero for all values of $n_\mathrm{cut}$, indicating that the models exhibit little systematic bias in their predictions.} The MAE decreases with increasing $n_\mathrm{cut}$, becoming as low as $1.24 \times 10^{-3}$ eV for $n_\mathrm{cut}$=15, illustrating the high predictive accuracy of the model.  \revA{Meanwhile, some predicted values exhibit relatively large deviations from the reference data. To assess the potential impact of these errors, we examine the probability density distributions of the prediction errors, as shown in Figs.~\ref{fig:probability_density}~(a-1)--(a-3). Note that the probability density is plotted on a logarithmic scale. All the distributions are sharply peaked around zero, consistent with the near-zero ME values in Fig.~\ref{fig:e111_prediction}. Furthermore, the probability of large prediction errors is extremely low. For instance, at $n_\mathrm{cut} = 15$, the probability of an absolute error exceeding 0.3 eV is only 0.007\%.} Consequently, these rare deviations have a negligible effect on the overall magnitude of $E_\mathrm{f}$, which is the primary quantity of interest in this study. Figs.~\ref{fig:e111_prediction}~(b-1)–(b-3) show the predicted values of $E^{i}_{[111]}$ for the SIA-containing strings. While the MAE also decreases with increasing $n_\mathrm{cut}$, the overall accuracy is lower than that for the SIA-free strings, e.g., the MAEs at $n_\mathrm{cut} = 10$ and 15 are an order of magnitude larger than those in the SIA-free string case. This difference likely arises from a much broader sampled range of $E^{i}_{[111]}$ for the SIA-containing strings, which spans from 0.54 to 8.42 eV due to the strong dependence of $E^{i}_{[111]}$ on the local SIA occupancy environment. \revA{Meanwhile, Figs.~\ref{fig:probability_density}~(b-1)--(b-3) show that, as in the SIA-free string case, the prediction error distributions are sharply peaked at zero, while the probability of large errors remains very low, e.g., at $n_\mathrm{cut} = 15$, only 0.036\% of the predictions have an absolute error exceeding 1.0 eV}. Considering this broad energy range \revA{and the low probability of large prediction errors}, the models can be regarded as sufficiently accurate in capturing the complex relationship between $E^{i}_{[111]}$ and local environment.

    \begin{figure*}[t]
    \centering
    \includegraphics[width=1.0\textwidth]{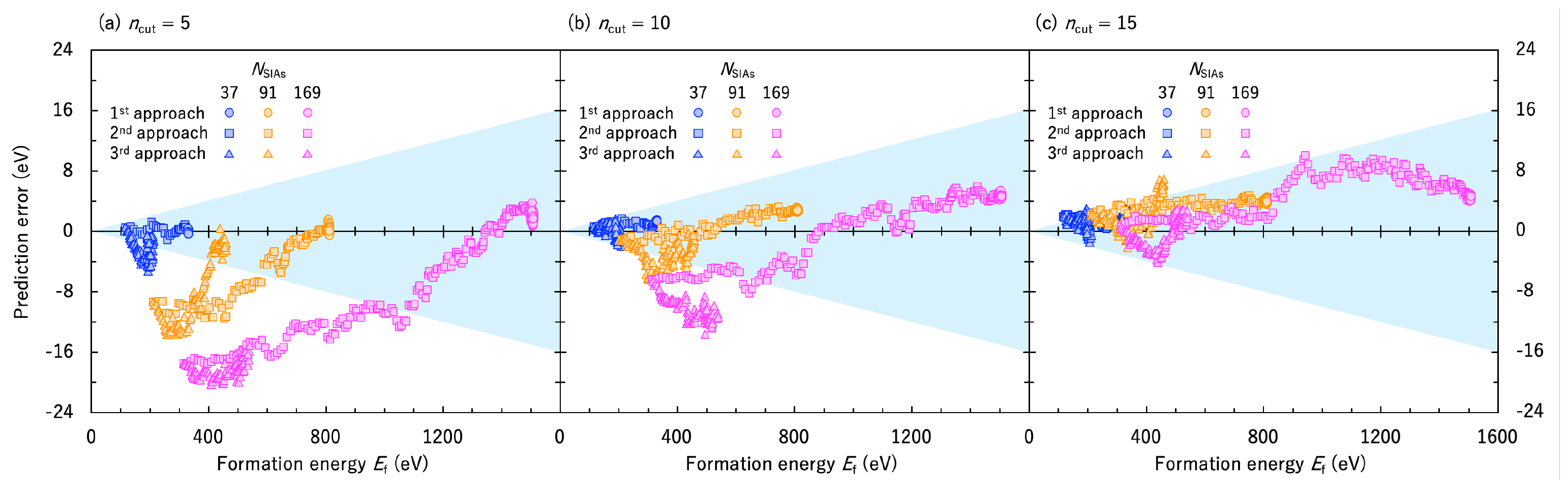}
    \caption{Prediction error for $E_\mathrm{f}$, evaluated from the predicted $E^{i}_{[111]}$ values using Eq.~(\ref{eq:2}), plotted as a function of $E_\mathrm{f}$ obtained from the system energy minimisation at $n_\mathrm{cut}$ = (a) 5, (b) 10, and (c) 15. The light-blue regions indicate the 1\% error range. \revA{The SIA configurations examined here are not included in either the prepared training or test datasets but are instead newly generated using the three approaches described in Sec.~II.}}
    \label{fig:ef_prediction}
    \end{figure*}

    \begin{figure}[htbp]
    \centering
    \includegraphics[width=1.0\linewidth]{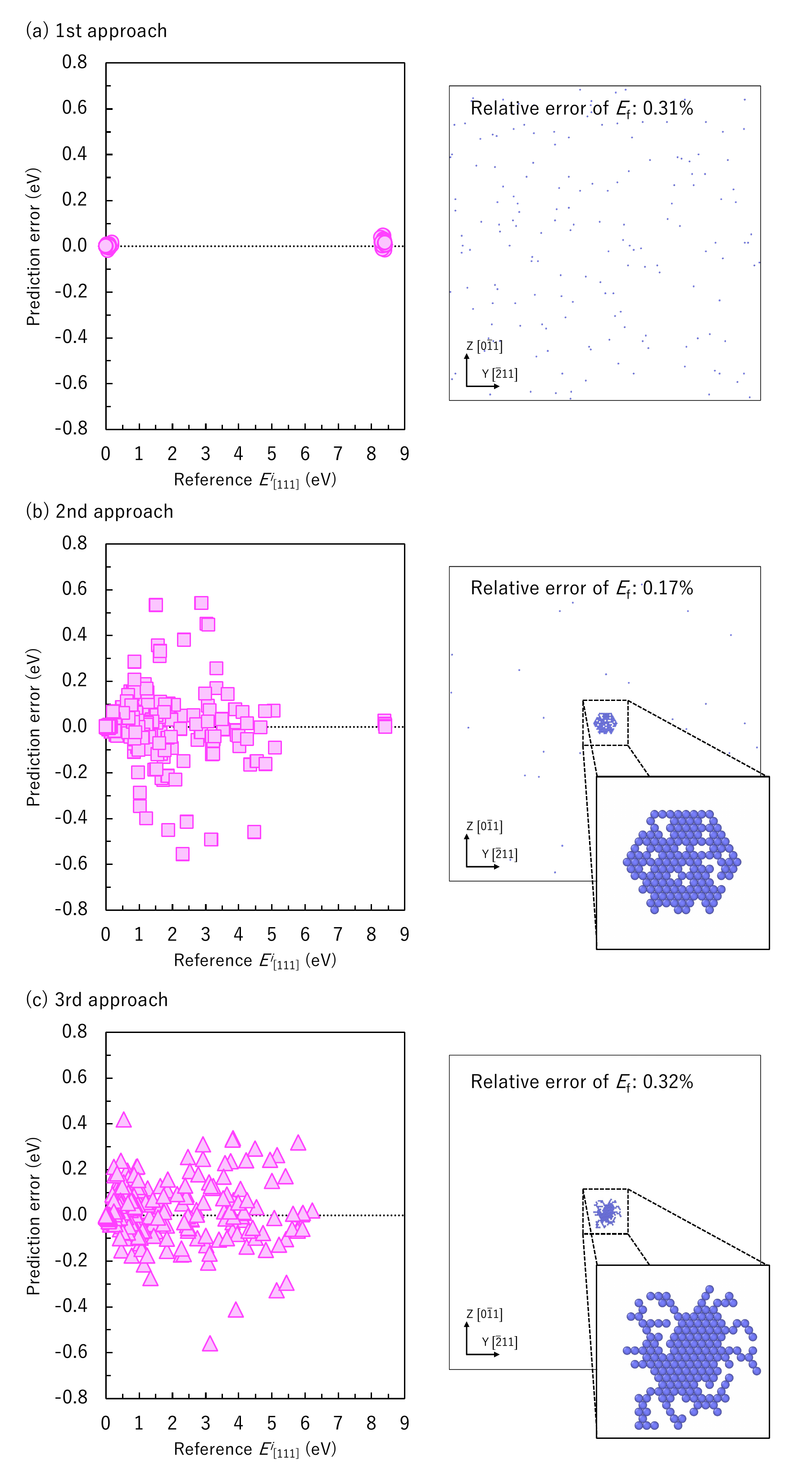}
    \caption{\revA{Decomposition of the error in $E_\mathrm{f}$ into the individual prediction errors in $E_\mathrm{[111]}$ for representative SIA configurations from Fig.~\ref{fig:ef_prediction} (c). The corresponding configuration is shown in each panel.}}
    \label{fig:ef_prediction_breakdown}
    \end{figure}
    
    \begin{figure}[htbp]
    \centering
    \includegraphics[width=1.0\linewidth]{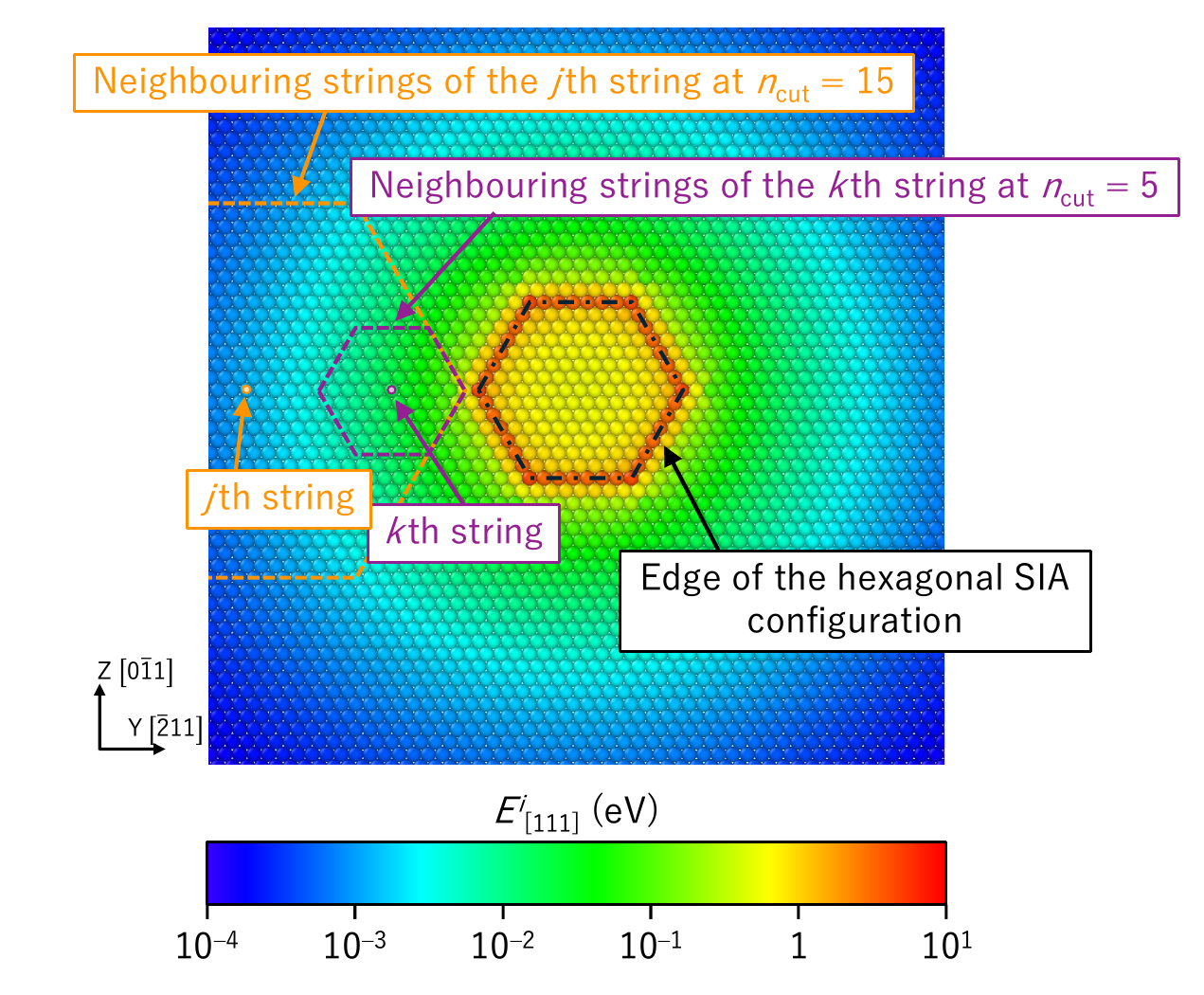}
    \caption{$E^{i}_{[111]}$ map for an SIA configuration with $N_{\mathrm{SIA}}$ = 169 in a hexagonal arrangement, calculated by minimizing the total energy of the system.}
    \label{fig:energy_map}
    \end{figure}
    
Using the models developed above, we predict the formation energies $E_\mathrm{f}$ of various SIA configurations with $N_{\mathrm{SIA}}$ = 37, 91, and 169 at different values of $n_\mathrm{cut}$, according to Eq.~(\ref{eq:2}). The prediction errors for these configurations are shown in Figs.~\ref{fig:ef_prediction}~(a)--(c) as a function of $E_\mathrm{f}$ obtained by the direct energy minimisation.  We note that the SIA configurations examined here are not included in either the training or test datasets. Instead, they are newly generated using the three approaches described in the previous section. 
\revA{This separate set of configurations is necessary because, in the training and test datasets prepared above, the individual $E_\mathrm{[111]}^i$ data points associated with a given SIA configuration are randomly assigned to either the training or test dataset. Consequently, the total $E_\mathrm{f}$ of a given configuration cannot be reconstructed solely from the data points in the test dataset. Therefore, to assess the model performance for $E_\mathrm{f}$ without using any data involved in model training, independent SIA configurations need to be generated and used exclusively for the $E_\mathrm{f}$ evaluation.}
At $n_\mathrm{cut}$ = 15, see Fig.~\ref{fig:ef_prediction}~(c), excellent predictive capacity is achieved across all the values of $N_{\mathrm{SIA}}$, with most configurations exhibiting errors within 1\%, as highlighted by the light-blue region in the figure. 
\revA{To clarify the origin of this high predictive accuracy, we decompose the error in $E_\mathrm{f}$ into the individual prediction errors in $E_\mathrm{[111]}$. Fig.~\ref{fig:ef_prediction_breakdown} presents this decomposition for one representative configuration for each approach of generating configurations, together with the corresponding SIA configurations. For the first approach (Fig.~\ref{fig:ef_prediction_breakdown}~(a)), the individual $E_\mathrm{[111]}^i$ prediction errors are extremely small, naturally resulting in a low $E_\mathrm{f}$ error of 0.31\%. By contrast, the errors for the second and third approaches (Figs.~\ref{fig:ef_prediction_breakdown}~(b) and (c)) exhibit considerably greater scatter. However, these errors are distributed around zero, consistent with the symmetric error distributions shown in Fig.~\ref{fig:probability_density}. Consequently, positive and negative contributions largely cancel out when summed over all strings, resulting in low $E_\mathrm{f}$ errors of 0.17\% and 0.32 \%, respectively. Thus, high predictive accuracy for $E_\mathrm{f}$ is maintained even for the geometrically complex SIA configurations shown in Figs.~\ref{fig:ef_prediction_breakdown}~(b) and (c).}

Meanwhile, the choice of $n_\mathrm{cut}$ has a significant effect on the accuracy of the prediction of $E_\mathrm{f}$. As shown in Fig.~\ref{fig:ef_prediction}~(b), when $n_\mathrm{cut}$ is reduced to 10, the predicted values of $E_\mathrm{f}$ for some configurations begin to show a sign of noticeable underestimation, deviating from the 1\% error range, particularly for the configurations with $N_{\mathrm{SIA}}$ = 91 and 169 generated using the third approach (triangle symbols) and those at lower values of $E_\mathrm{f}$ generated using the second approach (square symbols). The accuracy of predictions deteriorates further at $n_\mathrm{cut}$ = 5, where even configurations with $N_{\mathrm{SIA}}$ = 37 exhibit some significant underestimation  of energy (Fig.~\ref{fig:ef_prediction}~(a)). To investigate the fundamental origin of these underestimations, we create an $E^{i}_{[111]}$ map for an SIA configuration with $N_{\mathrm{SIA}}$ = 169 in a hexagonal arrangement, shown in  Fig.~\ref{fig:energy_map}. Note that the $E^{i}_{[111]}$ values shown here are calculated by a direct system energy minimisation rather than using the machine-learning models developed above. We find that $E^{i}_{[111]}$ decreases gradually as a function of distance from the edge of a hexagonal SIA configuration. For instance, the $j$th string, represented by an orange point, is 16 atomic distances from the hexagonal edge and has the energy of only 0.0013 eV. It should be noted that the $j$th string energy is predicted to be 0.0 eV by the models developed above when $n_\mathrm{cut} \le$ 15 because there are no SIA-containing strings within the neighbourhood range (orange hexagon). This discrepancy, however, has a negligible effect on the overall formation energy $E_\mathrm{f}$. In contrast, the $k$th string, represented by the purple point, is 6 atomic distances from the hexagonal edge and has the energy of 0.019 eV. Importantly, at $n_\mathrm{cut}$ = 5, this string's energy is entirely neglected by the models because no SIA-containing strings fall within the cutoff distance (purple hexagon), despite the fact that its energy is relatively high. The accumulation of such omissions results in a significant underestimation of $E_\mathrm{f}$ at low $n_\mathrm{cut}$, particularly for the configurations generated by the third approach, involving the investigation of fully-connected, no-hole SIA geometries, producing long-range strain fields. This explains the pronounced underestimation observed for the third approach at $n_\mathrm{cut} =$ 5 and 10, shown in Figs.~\ref{fig:ef_prediction}~(a) and (b), respectively. Meanwhile, the origin of the prediction error in the second approach can be interpreted as follows. As we follow the procedure described above, we note that initially the configurations have a perfect hexagonal arrangement. As more and more mono-SIAs are generated, the initial configuration gradually shrinks, increasing the overall $E_\mathrm{f}$. Since the range of strain fields produced by mono-SIAs has a relatively short range, as detailed for example by Eq.~(5) from Ref.~\cite{Warwick2025}, the underestimation of $E_\mathrm{f}$ caused by small $n_\mathrm{cut}$ is alleviated as the procedure progresses. Consequently, the prediction error decreases with increasing $E_\mathrm{f}$ for the configurations generated using the second approach at $n_\mathrm{cut}$ = 5 and 10, eventually falling within the 1\% error range.
    
These observations indicate that large, continuously connected configurations require a larger cutoff $n_\mathrm{cut}$, arising from the long-range strain fields, which in turn leads to a substantial computational cost of evaluating $E_\mathrm{f}$. One possible way of overcoming this limitation involves the incorporation of elastic field effects into the models. This would require evaluating the lattice distortion contribution to $E_\mathrm{f}$ using elasticity theory if an atomic string lies sufficiently far from the edges of the SIA configuration. Although an elasticity calculation would not necessarily provide accurate $E^{i}_{[111]}$ values for strings near the edges, it is valid for regions far from the edges, where the effect of the detailed configuration geometry is less significant. We note an earlier study, combining elasticity theory with atomistic modelling, which aimed to derive the core energy of a dislocation, where elasticity described the long-range elastic energy while atomistic simulations treated the core region \cite{Bertin2021PhysRevMaterials}.
\revA{Such a combination of configurational machine-learning and long-range elasticity could potentially reduce the cutoff required by the machine-learning model while retaining the long-range contribution to $E_\mathrm{f}$. It should be noted that its applicability, however, depends on whether the long-range strain field generated by a given SIA configuration can be adequately described by elasticity theory. The development and quantitative validation of such a coupled framework are left for future work.}


    \begin{figure*}[t]
    \centering
    \includegraphics[width=\textwidth]{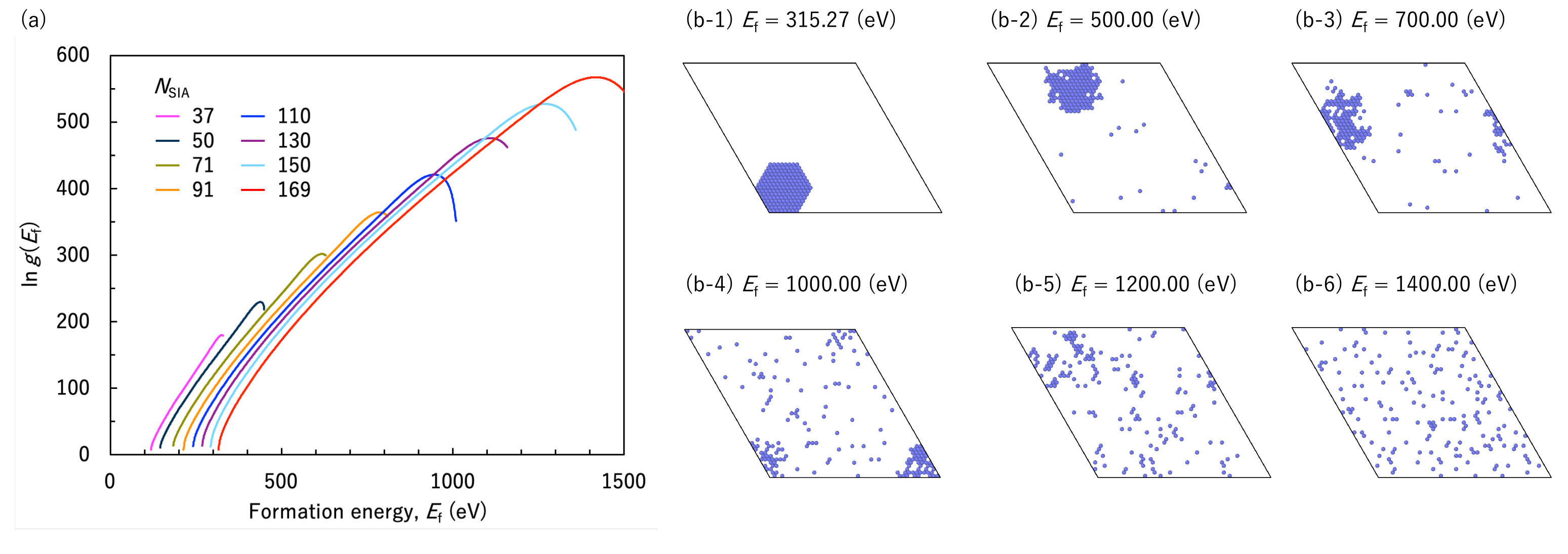}
    \caption{(a) Values of $g(E_\mathrm{f})$ as a function of $E_\mathrm{f}$, obtained from the Wang--Landau sampling for various values of $N_\mathrm{SIA}$. (b-1)--(b-6) Representative SIA configurations for $N_\mathrm{SIA} = 169$ at selected values of $E_\mathrm{f}$.}
    \label{fig:ln_g}
    \end{figure*}

\section{Density of microstates and thermodynamic properties of SIA configurations}

Using the predictive models for $E_\mathrm{f}$ developed above, we evaluate the density of configurational microstates for a system containing $N_\mathrm{SIA}$ SIAs.
Here, the density of microstates $g(E_\mathrm{f})$ denotes the number of distinct SIA configurations, corresponding to a given value of $E_\mathrm{f}$.
In statistical mechanics, $g(E_\mathrm{f})$ defines the canonical partition function \revA{$Z(T)$}, expressed as \cite{Kubo1965,McQuarrie2000}
\revA{
    \begin{equation}
    Z(T) = \sum_{E_\mathrm{f}} g(E_\mathrm{f}) e^{-E_\mathrm{f}/k_\mathrm{B} T} \text{,} \label{eq:partition_function}
    \end{equation}
}
where $T$ is the absolute temperature and $k_\mathrm{B}$ is the Boltzmann constant. Once \revA{$Z(T)$} is computed, the various thermodynamic quantities such as the free energy, mean energy, and entropy can be directly evaluated as functions of temperature. These provide essential insight into the thermodynamic stability of SIA configurations associated with a given $N_\mathrm{SIA}$. To the best of our knowledge, no study to date has attempted to determine $g(E_\mathrm{f})$ for SIA configurations, despite its evident significance for applications.

To compute $g(E_\mathrm{f})$, we employ the Wang--Landau sampling method, which is a Monte Carlo technique designed to explore the energy space uniformly \cite{Wang2001PhysRevLett,Wang2001PhysRevE,Landau2004BrazJPhys}. In this algorithm, a random walk is performed in the configurational space with transition probabilities chosen to be inversely proportional to the current estimate of $g(E_\mathrm{f})$. As a result, a flat histogram in energy space is gradually established, while $g(E_\mathrm{f})$ is refined iteratively throughout the simulation. In particular, here we use a multi-range Wang--Landau sampling, where multiple random walks are performed over different energy ranges and the resulting $g(E_\mathrm{f})$ values are subsequently merged across the adjacent ranges \cite{Wang2001PhysRevE}. This approach accelerates the exploration of the full energy space, enabling an efficient parallelisation of calculations for independent energy ranges. We consider a two-dimensional parallelogram cell containing $L \times L$ lattice sites with periodic boundary conditions, with basis vectors  $\boldsymbol{e}_1$ and $\boldsymbol{e}_2$ given by Eqs.~(\ref{eq:e4}) and (\ref{eq:e5}), respectively. The value of $L$ is set to 45 (see Section~IV and Fig.~S2 of the Supplementary Materials for a detailed discussion of the role played by $L$). The total of $N_\mathrm{SIA}$ SIAs are initially placed at arbitrary positions, and at each Monte Carlo step, a randomly chosen SIA is displaced to a random SIA-free location. We use the machine learning models developed above for $n_\mathrm{cut} = 15$ to evaluate $E_\mathrm{f}$ for each configuration. The energy ranges considered here are listed in Table S1 in the Supplementary Materials, and the width of an energy bin for constructing the histogram is set to 1 eV. Note that we do not evaluate $g(E_\mathrm{f})$ in the very high $E_\mathrm{f}$ region, as the computational cost in this region is high due to the strong variation of $g(E_\mathrm{f})$. Nevertheless, this exclusion does not affect the analysis below as the contribution from the high energy states to the canonical partition function \revA{$Z(T)$} are negligible owing to their suppression by the exponential Boltzmann temperature factor in Eq.~(\ref{eq:partition_function}). The simulation starts with an initial modification factor $f = f_0 = e^1$, and $g(E_\mathrm{f})$ is updated by multiplying the current value of $g(E_\mathrm{f})$ by $f$ each time a state with $E_\mathrm{f}$ is visited. The histogram flatness criterion is set to 80\%, and once the criterion is met, the $f$ is reduced according to $f \rightarrow \sqrt{f}$. This procedure is repeated until $f < e^{10^{-8}}$, at which point the $g(E_\mathrm{f})$ is considered converged. After $g(E_\mathrm{f})$ is obtained for each energy range, the results from the neighbouring ranges are merged using the procedure described in Sec.~III of Supplementary Materials.

Fig.~\ref{fig:ln_g}~(a) shows the resulting $g(E_\mathrm{f})$ plotted as a function of $E_\mathrm{f}$ for various $N_\mathrm{SIA}$. A general trend is observed regardless of the choice of $N_\mathrm{SIA}$: except in the high-$E_\mathrm{f}$ region, $g(E_\mathrm{f})$ increases rapidly as a function of $E_\mathrm{f}$, accounting for the progressively more disordered configurations, see Figs.~\ref{fig:ln_g}~(b-1)--(b-6). At sufficiently large $E_\mathrm{f}$, $g(E_\mathrm{f})$ reaches a peak and then decreases. This decreasing behaviour can be understood as follows. The value of $E_\mathrm{f}$ attains its maximum when the sum of interaction energies between the SIAs is minimised. Configurations satisfying this condition are those where each SIA is separated from the other defects as far apart as possible given the constraint imposed by the average spatial density of defects. Since this requirement imposes a strong geometric constraint, the number of accessible configurations is severely restricted; in practice, only a small fraction of states attain such high values of $E_\mathrm{f}$. Consequently, $g(E_\mathrm{f})$ decreases as $E_\mathrm{f}$ approaches its maximum value at the high $E_\mathrm{f}$ region. A similar behaviour of $g(E_\mathrm{f})$ exhibiting a peak has also been found for the Ising model \cite{Wang2001PhysRevE}.

    \begin{figure*}[t]
    \centering
    \includegraphics[width=\textwidth]{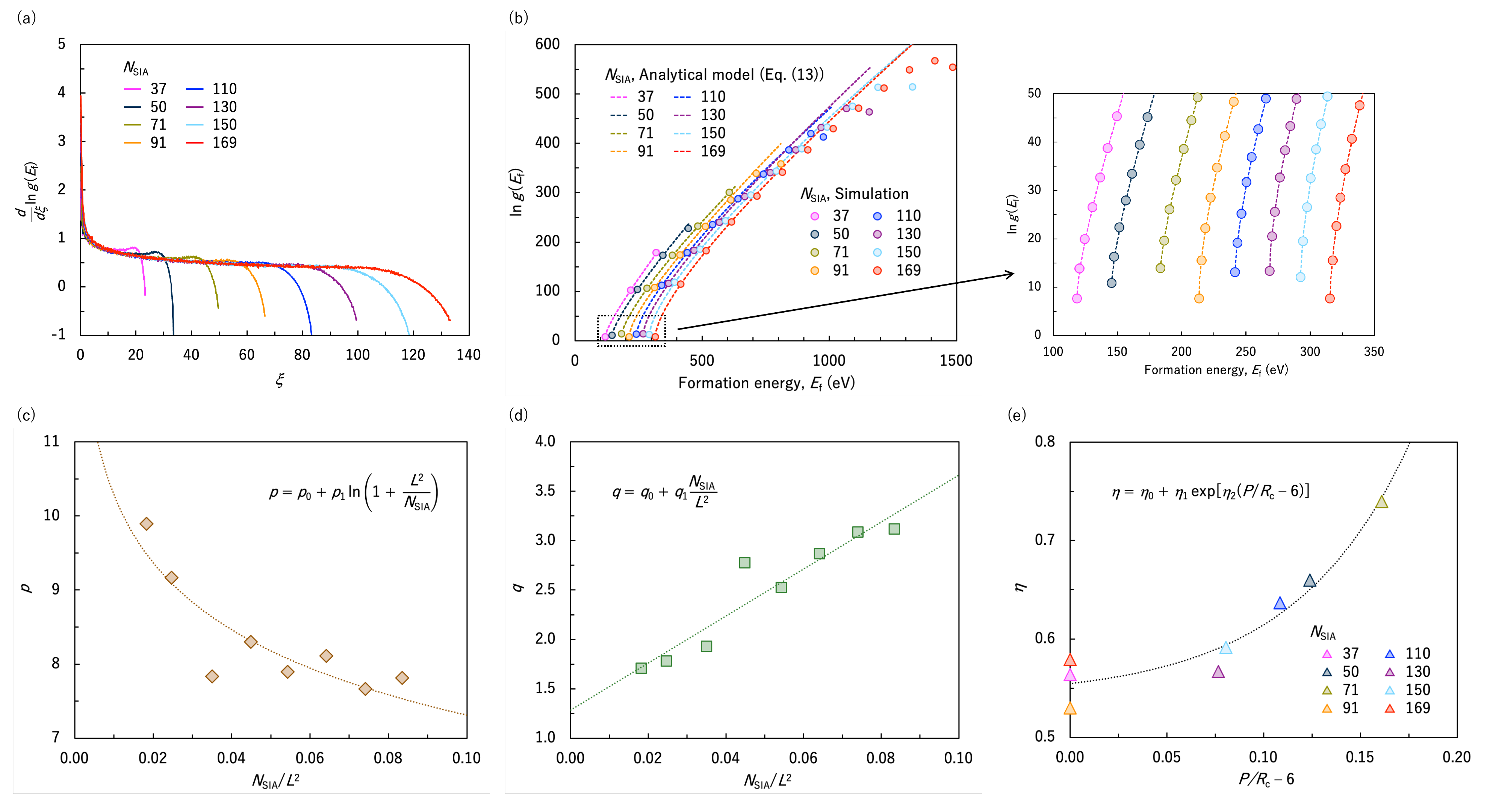}
    \caption{(a) Slopes of the curves of $\ln g(E_\mathrm{f})$ as a function of $\xi$ for different values of $N_\mathrm{SIA}$. (b) Values of $\ln g(E_\mathrm{f})$ predicted by the analytical model in Eq.~(\ref{eq:113}) (dashed curves), compared with selected simulation data (circular symbols). (c) Fitted values of $p$ as a function of $N_\mathrm{SIA}/L^2$. The dotted curve represents the analytical expression given by Eq.~(\ref{eq:14}). (d) Fitted values of $q$ as a function of $N_\mathrm{SIA}/L^2$. The dotted line shows the linear fit described by Eq.~(\ref{eq:15}). (e) Fitted values of $\eta$ as a function of $P/R_\mathrm{c} - 6$. The dotted curve represents the analytical expression given by Eq.~(\ref{eq:16}).}
    \label{fig:ln_g_fitting}
    \end{figure*}

We now consider an analytical form for $\ln g(E_\mathrm{f})$ as a function of $E_\mathrm{f}$, which enables efficient evaluation of $\ln g(E_\mathrm{f})$ for arbitrary $N_\mathrm{SIA}$. Fig.~\ref{fig:ln_g_fitting}~(a) shows the slopes of curves of $ \ln g(E_\mathrm{f})$, shown in Fig.~\ref{fig:ln_g}~(a), as a function of the dimensionless variable $\xi$, defined by
    \begin{equation}
    \xi = \frac{E_\mathrm{f} - E_\mathrm{f,ground}}{E_\mathrm{f}^\text{mono-SIA}} , \label{eq:11}
    \end{equation}
where $E_\mathrm{f}^\mathrm{mono-SIA}$ denotes the $E_\mathrm{f}$ of an isolated mono-SIA with a $\left< 111 \right>$ crowdion configuration, and $E_\mathrm{f,ground}$ is the $E_\mathrm{f}$ of the ground-state configuration for each $N_\mathrm{SIA}$. As a general trend, the slope of the  $\ln g(E_\mathrm{f})$ plot varies distinctly over three characteristic regions, correlated with the range of $\xi$. If $\xi$ is very small, i.e., $E_\mathrm{f}$ is close to $E_\mathrm{f,ground}$, the slope is initially high, but flattening rapidly as $\xi$ increases. This is followed by a plateau region in the intermediate $\xi$ range, indicating an almost linear increase of $\ln g(E_\mathrm{f})$ with respect to $E_\mathrm{f}$. Eventually, the slope decreases in the limit where $\xi$ is close to its maximum value. Since the values of $\ln g(E_\mathrm{f})$ in the low to intermediate $\xi$ range are particularly important for practical applications (e.g. for the evaluation of the partition function \revA{$Z(T)$}), we focus our attention on this range and introduce the following analytical form for the slope of the curve:
    \begin{equation}
    \frac{d}{d \xi} \ln g(E_\mathrm{f}) = p - r \eta + \frac{q r^{\eta}}{\xi^{1-\eta}} \text{,} \label{eq:12}
    \end{equation}
where $p, q$, $r$, and $\eta$ are fitting parameters, satisfying conditions $p > r\eta$, $q > 0$, $r > 0$, and $0 < \eta < 1$. In this expression, the term $p - r\eta$ corresponds to the plateau value of the slope, while the last term accounts for its rapid increase as $\xi$ approaches zero \revA{(see Sec.~VI of the Supplementary Materials for further details)}. An analytical expression for $\ln g(E_\mathrm{f})$ is then obtained by integrating Eq.~(\ref{eq:12}) with respect to $\xi$:
    \begin{equation}
        \ln g(E_\mathrm{f}) = \left( p - r \eta \right) \xi + \frac{q r^{\eta}}{\eta}  \xi^{\eta} + C \text{,} \label{eq:113}
    \end{equation}
where $C = \ln g(E_\mathrm{f,ground})$, which needs to be specified for each value of $N_\mathrm{SIA}$. The dashed curves in Fig.~\ref{fig:ln_g_fitting}~(b) show the predictions derived from Eq.~(\ref{eq:113}), with parameters $p, q$, and $\eta$ fitted to the simulation results shown in Fig.~\ref{fig:ln_g}~(a). Here, $r$ is fixed to the numerical value of $E_\mathrm{f}^\text{mono-SIA}$ for all $N_\mathrm{SIA}$ because it is found to yield good fitting results. Some selected simulation data are also shown as circular symbols for comparison. The analytical expression reproduces the simulation results very well, except in the high $E_\mathrm{f}$ region, which is excluded from the present analysis. \revA{For a quantitative assessment of the fitting quality, we evaluate the MAE of the predicted $\ln g (E_\mathrm{f})$ relative to the simulation data over the low to intermediate $\xi$ range. The MAE remains below 1 for all $N_\mathrm{SIA}$; the corresponding values are summarised in Table \hl{S2} of the Supplementary Materials.}

Next, to establish a systematic procedure for determining the parameter values for any given $N_\mathrm{SIA}$, we examine the $N_\mathrm{SIA}$ dependence of the fitted values of $p$, $q$, and $\eta$. As shown in Fig.~\ref{fig:ln_g_fitting}~(c),  parameter $p$ is found to follow an empirical relation:
    \begin{equation}
    p = p_0 + p_1 \ln \left(1 + \frac{L^2}{N_\mathrm{SIA}} \right) \text{,} \label{eq:14}
    \end{equation}
where $p_0$ and $p_1$ are fitted to be 4.094 and 1.342, respectively, and $L$ denotes the number of lattice sites along each simulation cell axis, as mentioned above (see also Sec.~IV in Supplementary Materials for the variation of the parameters at different $L$ values). While some data points deviate from the fitted curve---particularly for $N_\mathrm{SIA} = 71$ (i.e., $N_\mathrm{SIA}/L^2 = 0.035$)---the overall agreement is excellent, indicating a systematic dependence of $p$ on $N_\mathrm{SIA}/L^2$, i.e., the SIA concentration in the system. Furthermore, a linear dependence of $q$ on $N_\mathrm{SIA}/L^2$ is observed, as shown in Fig.~\ref{fig:ln_g_fitting}~(d):
    \begin{equation}
    q = q_0 + q_1 \frac{N_\mathrm{SIA}}{L^2} \text{,} \label{eq:15}
    \end{equation}
where $q_0$ and $q_1$ are fitted to be 1.287 and 23.778, respectively. On the other hand, we find that parameter $\eta$ does not exhibit a simple dependence on $N_\mathrm{SIA}/L^2$. Rather, as shown in Fig.~\ref{fig:ln_g_fitting}~(e), $\eta$ is correlated with the ratio $P/R_\mathrm{c}$, where $P$ denotes the perimeter of the ground-state configuration and $R_\mathrm{c}$ is its characteristic radius, which is uniquely defined for a given $N_\mathrm{SIA}$ (see Sec.~V and Fig.~S3 in Supplementary Materials for detailed definitions). Accordingly, the ratio of $P/R_\mathrm{c}$ represents the perimeter normalised by the characteristic radius. The relationship between $\eta$ and $P/R_\mathrm{c}$ can be well approximated by the relation
    \begin{equation}
    \eta = \eta_0 + \eta_1 \exp \left[ \eta_2 \left( \frac{P}{R_\mathrm{c}} - 6 \right) \right] \text{,} \label{eq:16}
    \end{equation}
where $\eta_0$, $\eta_1$, and $\eta_2$ are fitted to be 0.541, 0.0140, and 16.619, respectively. The meaning of $\eta$ can be understood as follows. Since $P/R_\mathrm{c} = 6$ for a perfectly hexagonal configuration, regardless of its size, the quantity $P/R_\mathrm{c} - 6$ effectively measures the deviation of the ground-state configuration from a perfect hexagonal geometry. Eq.~(\ref{eq:16}) therefore shows that $\eta$ can be interpreted as an effective size-independent parameter characterising the morphological irregularity of the ground-state configuration of an SIA cluster. As demonstrated below, it is this parameter $\eta$ that critically controls both the thermodynamic properties and dynamics of loops. It should also be highlighted that the distinct dependences of $p, q$, and $\eta$ confirm that the formulation adopted in Eq.~(\ref{eq:113}) effectively decouples two independent contributions: the ground-state configurational characteristics, described by $\eta$, and the defect density effect, reflected in parameters $p$ and $q$.
\revA{The validation of the functional forms chosen for $p$ (Eq.~(\ref{eq:14})), $q$ (Eq.~(\ref{eq:15})), and $\eta$ (Eq.~(\ref{eq:16})), together with analysis of their fit quality and sensitivity, is discussed in Secs.~VII and VIII of the Supplementary Materials.}

    \begin{figure}[htbp]
    \centering
    \includegraphics[width=\linewidth]{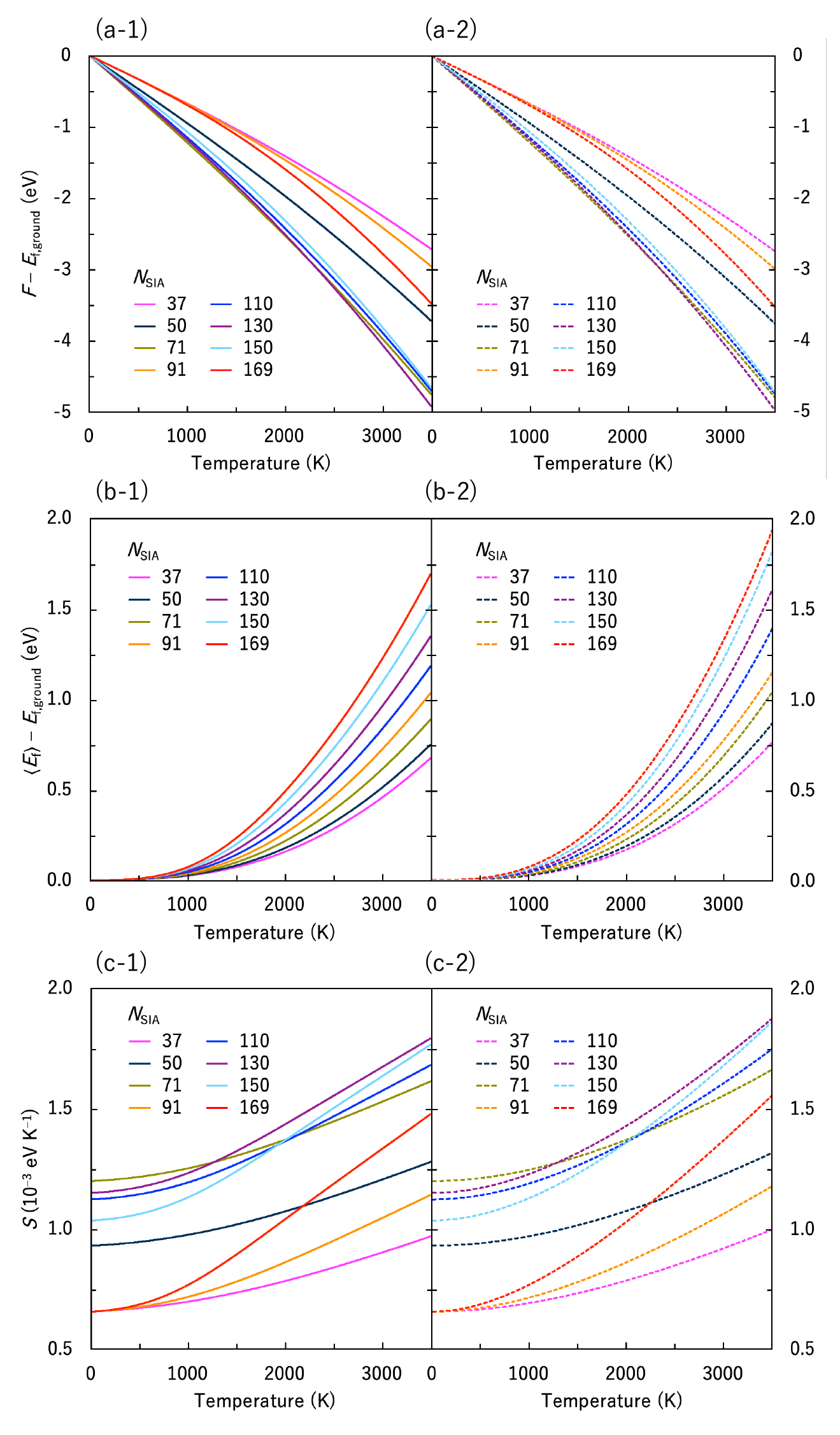}
    \caption{Temperature dependence of (a) $F$, (b) $\langle E_\mathrm{f} \rangle$, and (c) $S$ for various values of $N_\mathrm{SIA}$. The values for (a-1), (b-1) and (c-1) are calculated from Eqs.~(\ref{eq:17}), (\ref{eq:18}), and (\ref{eq:19}), respectively, while those for (a-2), (b-2) and (c-2) are predicted by Eqs.~(\ref{eq:20}), (\ref{eq:21}), and (\ref{eq:22}), respectively.}
    \label{fig:F_S_Emean}
    \end{figure}

Importantly, the analytical model for $\ln g(E_\mathrm{f})$ proposed here provides a numerically convenient and efficient framework for investigating the thermodynamic properties of SIA configurations for any density of SIA defects in a material. For each $N_\mathrm{SIA}$, the model requires only $P$ in Eq.~(\ref{eq:16}), together with $E_\mathrm{f,ground}$ and $C = \ln g(E_\mathrm{f,ground})$ in Eqs.~(\ref{eq:11}) and (\ref{eq:113}). These values are fully determined by the information derived from the ground-state properties of the system. This means that, once this information is known, the model enables predicting $\ln g(E_\mathrm{f})$ over the practically relevant energy range and constructing the partition function \revA{$Z(T)$} according to Eq.~(\ref{eq:partition_function}).
This, in turn, enables the evaluation of key thermodynamic quantities as functions of temperature, thereby providing a fundamental understanding of the stability of SIA configurations.

To demonstrate the usefulness of this framework, using the proposed analytical model Eq.~(\ref{eq:113}) together with Eqs.~(\ref{eq:14})--(\ref{eq:16}), we evaluate \revA{$Z(T)$} and then compute the configurational free energy $F$, the mean energy $\langle E_\mathrm{f} \rangle$, and entropy $S$ as functions of temperature for the various values of $N_\mathrm{SIA}$. The calculations follow the statistical mechanics analysis \cite{Kubo1965,McQuarrie2000} 
\revA{
    \begin{align}
    F &= -k_\mathrm{B} T\ln{Z(T)} \text{,} \label{eq:17} \\
    \langle E_\mathrm{f} \rangle &= \frac{1}{Z(T)}\sum_{E_\mathrm{f}} E_\mathrm{f} g(E_\mathrm{f}) e^{-E_\mathrm{f}/k_\mathrm{B} T} \text{,} \label{eq:18} \\
    S &= \frac{1}{T} ( \langle E_\mathrm{f} \rangle - F ) \text{,} \label{eq:19}
    \end{align}
}
and the results are illustrated in Figs.~\ref{fig:F_S_Emean}~(a-1), (b-1), and (c-1), respectively. As shown in Fig.~\ref{fig:F_S_Emean}~(b-1), $\langle E_\mathrm{f} \rangle$ exhibits a monotonic dependence on $N_\mathrm{SIA}$; specifically, a larger $N_\mathrm{SIA}$ leads to a higher $\langle E_\mathrm{f} \rangle - E_\mathrm{f,ground}$. In contrast, as shown in Figs.~\ref{fig:F_S_Emean}~(a-1) and (c-1), no simple interpretable dependence of $F$ and $S$ on $N_\mathrm{SIA}$ is observed. For instance, at low and intermediate temperatures, smaller $S$ values are observed for $N_\mathrm{SIA} = $\,37, 91, and 169. In these `magic SIA number' cases, the ground state SIA cluster configurations form perfect hexagons, yielding a single configurational pattern, invariant (and degenerate) only with respect to lattice translations. Consequently, the value of $S$ at $T = 0$ is the lowest among all $N_\mathrm{SIA}$ values, leading to smaller $S$ and correspondingly weaker temperature dependences of $F$ at low and intermediate temperatures. This observation correlates well with a study of dislocation line orientation-dependent core energies, exhibiting a non-monotonic trend in the dislocation core energy as a function of the orientation of the line with respect to the orientation of the Burgers vector \cite{Bertin2021PhysRevMaterials}. The core energy of a dislocation exhibits non-analytic variation, reflecting the presence of irregularities in the core structure along the dislocation line, and highlighting the significance of incorporating detailed geometric features of dislocation line structure into models for defect configurations and defect dynamics.

    \begin{figure}[htbp]
    \centering
    \includegraphics[width=0.9\linewidth]{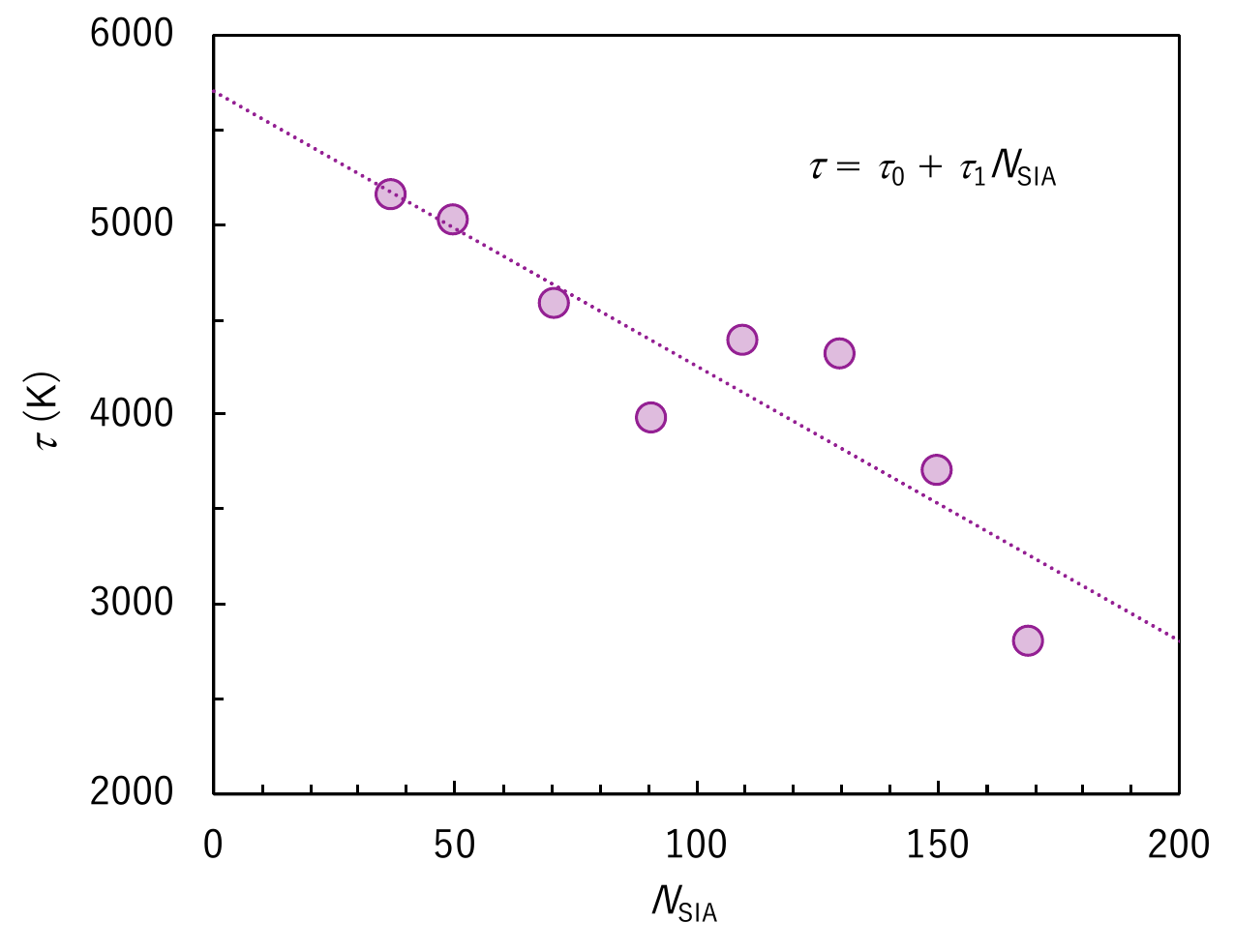}
    \caption{\revA{Dependence of $\tau$ on $N_\mathrm{SIA}$. The dotted line illustrates a linear fit, $\tau = \tau_0 + \tau_1 N_\mathrm{SIA}$, with fitted values of $\tau_0 = 5070.722$ K and $\tau_1 = -14.541$ K.}}
    \label{fig:alpha_tau}
    \end{figure}

To quantitatively demonstrate the influence of configurational variations of loops on their thermodynamic properties, we derive an analytical expression for the temperature dependence of $F$ using the parameter $\eta$:
    \revA{\begin{align}
    F &= k_\mathrm{B} \ln{g(E_\mathrm{f,ground})} \left[ \tau \eta \exp \left( -\frac{T}{\tau \eta} \right) - \frac{T^2}{2 \tau \eta} - \tau \eta \right] \nonumber \\
    &\quad + E_\mathrm{f,ground} \text{.} \label{eq:20}
    \end{align}}
Here, \revA{$\tau$ is a fitting parameter}, while $\eta$ characterises morphological deviations of the ground-state configuration of an SIA cluster from ideal hexagonal geometry, as discussed above \revA{(see Sec.~\hl{IX} of the Supplementary Materials for a detailed derivation of the analytical expression)}. Fig.~\ref{fig:F_S_Emean}~(a-2) shows the predicted temperature dependence of $F$ for each $N_\mathrm{SIA}$, with \revA{the parameter $\tau$} fitted to the corresponding data in Fig.~\ref{fig:F_S_Emean}~(a-1). Very good agreement is found across the entire temperature range considered in the study. Furthermore, \revA{the $\tau$} appears to vary linearly as a function of  $N_\mathrm{SIA}$, as shown in Fig.~\ref{fig:alpha_tau}. These results indicate that Eq.~(\ref{eq:20}) effectively separates $F$ into ground-state configurational contributions, characterised by $\eta$, and the size-dependent terms, characterised by \revA{the parameter $\tau$}.

Since $F$ serves as a generating function for thermodynamic quantities \cite{Kubo1965,McQuarrie2000}, Eq.~(\ref{eq:20}) enables deriving the temperature dependence of $\langle E_\mathrm{f} \rangle$ and $S$ in analytical form as:
    \revA{\begin{align}
    \langle E_\mathrm{f} \rangle &= F - T \frac{\partial F}{\partial T}  \nonumber\\
    &= k_\mathrm{B} \ln g(E_\mathrm{f,ground}) \bigg[ (\tau \eta + T) \exp \left( - \frac{T}{\tau \eta} \right)  \nonumber \\
    &\quad + \frac{T^2}{2 \tau \eta} - \tau \eta \bigg] + E_\mathrm{f,ground} \text{,} \label{eq:21} \\[6pt]
    S &= -\frac{\partial F}{\partial T} \nonumber \\
    &= k_\mathrm{B} \ln g(E_\mathrm{f,ground}) \left[ \exp \left( - \frac{T}{\tau \eta} \right) + \frac{T}{\tau \eta} \right] \text{.} \label{eq:22}
    \end{align}}
The dashed curves in Figs.~\ref{fig:F_S_Emean}~(b-2) and (c-2) show the values of $\langle E_\mathrm{f} \rangle$ and $S$ predicted by these analytical expressions. The overall agreement with the data in Figs.~\ref{fig:F_S_Emean}~(b-1) and (c-1) is excellent. It should be emphasised that \revA{the value of parameter $\tau$ is} determined solely by fitting to $F$, without any adjustment of values to $\langle E_\mathrm{f} \rangle$ or $S$. Taken together, Eqs.~(\ref{eq:20})--(\ref{eq:22}) establish a quantitative link between the geometry of a loop and its thermodynamic properties through parameter $\eta$.


\section{Loop dynamics through self-climb}

In this section, we examine the influence of configurational variations in loop geometry on its dynamics. In particular, we focus exclusively on self-climb-driven loop dynamics, rather than glide, because our atomistic simulations show that variations in loop shape have little effect on glide behaviour (see Sec.~X and Fig.~S6 in Supplementary Materials). Experimental studies have confirmed that self-climb (also referred to as conservative climb) plays a crucial role in loop dynamics \cite{Turnbull1970PhilosMag,Yao2010PhilosMag,Swinburne2016SciRep}. This process occurs through fluctuations in the loop geometry, mediated by cumulative pipe-diffusion of constituent SIAs along the loop perimeter. Importantly, it does not involve the emission or absorption of point defects by loops. Several models have been developed to describe this process, including analytical approaches \cite{Kroupa1961PhilosMag1,Kroupa1961PhilosMag2}, atomistic kinetic Monte Carlo (kMC) simulations \cite{Swinburne2016SciRep,Okita2016ActaMater,Hayakawa2016NuclMaterEnergy9-592,Hayakawa2018PhilosMag}, and discrete dislocation dynamics \cite{Niu2019IntJPlast}. Furthermore, computational studies have indicated that variations of loop geometric configurations represent a key stage in the self-climb process, determining the self-climb mobility \cite{Swinburne2016SciRep,Okita2016ActaMater}. Still, a quantitative link between the climb-mediated loop dynamics and configurational variations associated with loop geometry remains poorly understood and has only rarely been addressed in literature. This fundamental aspect of dislocation loop dynamics has remained one of the critical limitations in the context of a quantitative framework for modelling microstructural evolution under irradiation conditions.

We perform atomistic kMC simulations to investigate the behaviour of a dislocation loop undergoing self-climb, with a particular emphasis on the effect of configurational variations in the loop geometry. We simulate the individual events, mediating the pipe-diffusion of the SIAs forming the loop. These stochastic diffusion events induce fluctuations of the loop geometry and ultimately result in the Brownian motion of the loop's centre of mass. We consider a two-dimensional coordinate system spanned by the basis vectors $\boldsymbol{e}_1$ and $\boldsymbol{e}_2$, as defined in Eqs.~(\ref{eq:e4}) and (\ref{eq:e5}), respectively, \revA{with SIA positions restricted to the lattice sites (see also Fig.~\ref{fig:model_illustration}~(c)). Each lattice site has six nearest-neighbour sites, and a single kMC event corresponds to the hop of one SIA to one of these sites. Accordingly, the hopping vector $\boldsymbol{h}$ is given by
    \begin{equation}
    \boldsymbol{h} \in \left\{ \pm\boldsymbol{e}_1, \, \pm\boldsymbol{e}_2, \, \pm\boldsymbol{e}_3 \right\} \text{,}
    \end{equation}
where $\boldsymbol{e}_3$ is the third symmetry-related nearest-neighbour vector defined in Eq.~(\ref{eq:e6}). Hops to sites already occupied by another SIA are excluded. In addition, the allowed hopping events are restricted so as to preserve full connectivity of the SIA configuration; specifically, any hop that would leave an SIA isolated from all other SIAs is prohibited.} A total of $N_\mathrm{SIA}$ SIAs are initially positioned at the lattice sites corresponding to the ground state configuration, derived from the Wang--Landau sampling. \revA{We do not consider SIAs entering the loop from the bulk or leaving the loop, nor vacancy absorption by the loop. Consequently, the value of $N_\mathrm{SIA}$ is conserved throughout the simulations.} At each kMC step, the event list is constructed from all the possible patterns of SIA hopping processes under the current SIA configuration. Following the Kang--Weinberg model \cite{Kang1989JChemPhys}, we evaluate the rate constant $k$ for each event as
    \begin{equation}
    k = \nu_0 \exp{\left( - \frac{E_\mathrm{m} + \Delta E_\mathrm{f}/2}{k_\mathrm{B} T} \right)} \text{,} \label{eq:23}
    \end{equation}
where $\nu_0 = 10^{13}$ Hz is the attempt frequency and $E_\mathrm{m} = 2.359$ eV is the migration barrier of SIA pipe-diffusion along the loop edge in the absence of a change in $E_\mathrm{f}$ before and after the event. The latter is obtained from a climbing-image nudged elastic band calculation \cite{Henkelman2000JChemPhys} (see Sec.~XI and Fig.~S7 in Supplementary Materials for details). Parameter $\Delta E_\mathrm{f}$ denotes the difference in $E_\mathrm{f}$ before and after the event, which is evaluated using the machine-learning models for $n_\mathrm{cut} = 15$ developed in this work. From the trajectory of the loop centre of mass, we compute the mean squared displacement to obtain the diffusion constant. Following the approach described in Refs. \cite{Du2012PhysRevB,Ramasubramaniam2008JMaterSci}, the trajectory is divided into segments, and the diffusion constant $D$ is obtained as the time-weighted average of diffusion constants computed for each segment $i$:
    \begin{align}
    D &= \sum_i D_i \frac{\Delta t_i}{t_\mathrm{tot}} \text{,} \label{eq:24} \\
    D_i &= \frac{[ \boldsymbol{r}(t_i) - \boldsymbol{r}(t_{i-1}) ]^2}{2d \Delta t_i} \text{,} \label{eq:25}
    \end{align}
where $\Delta t_i = t_i - t_{i-1}$ is the duration of segment $i$, $t_\mathrm{tot}$ is the total simulation time, $\boldsymbol{r}(t)$ denotes the position of the loop's centre of mass at time $t$, and $d = 2$ is the dimensionality of the system.

    \begin{figure}[t]
    \centering
    \includegraphics[width=0.9\linewidth]{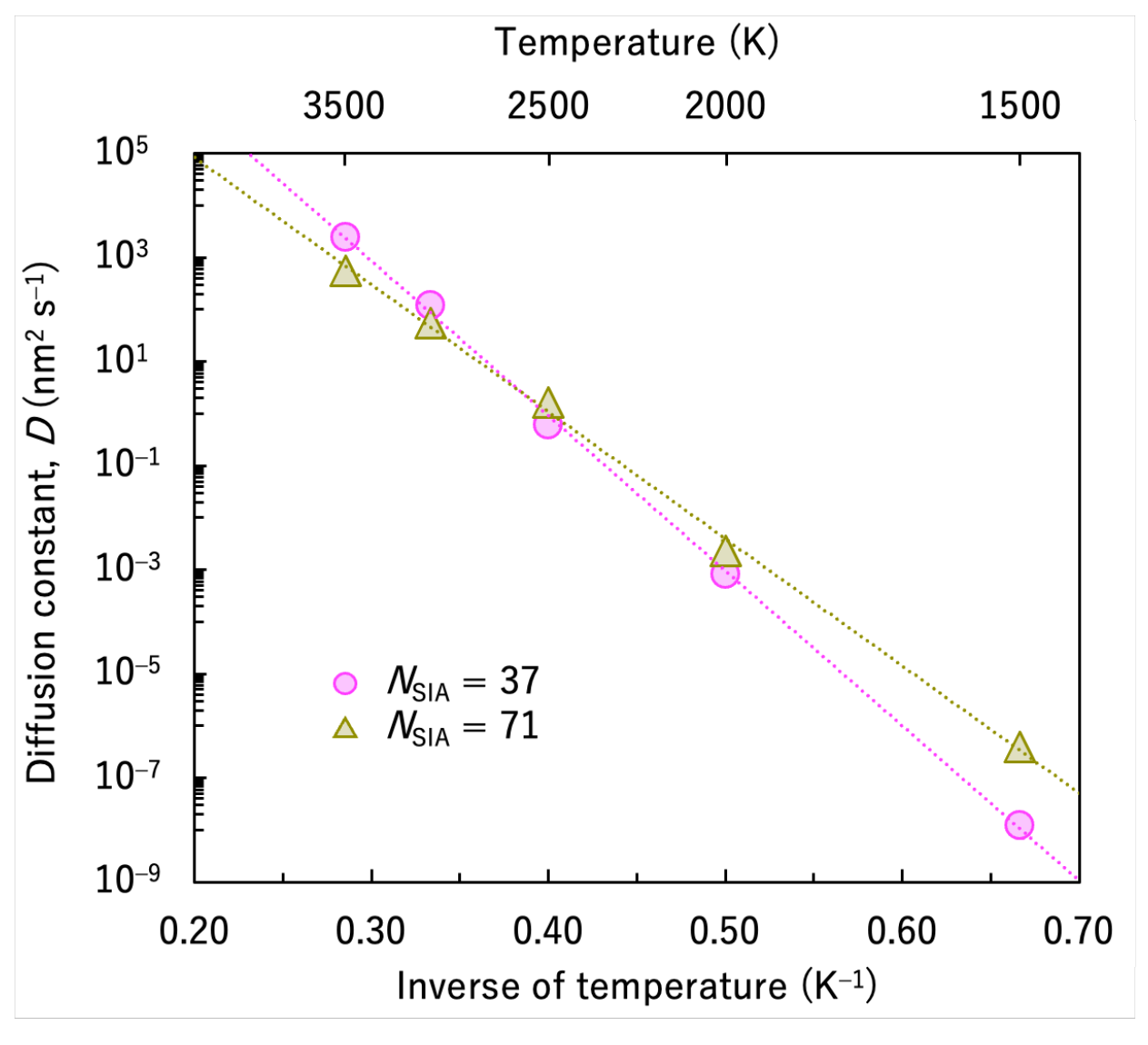}
    \caption{Temperature dependence of $D$ for $N_\mathrm{SIA} =$ 37 and 71.}
    \label{fig:self-climb_D}
    \end{figure}

\begin{figure*}[t]
    \centering
    \includegraphics[width=0.7\linewidth]{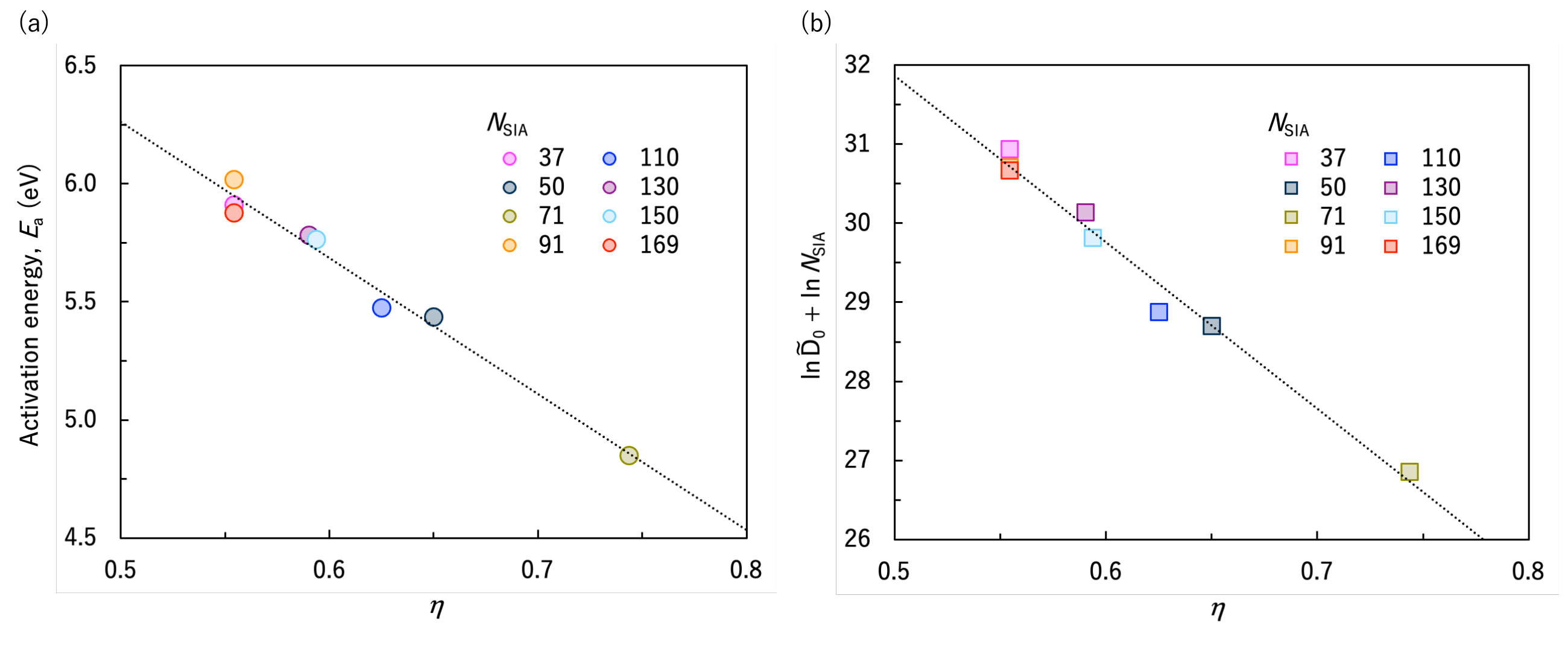}
    \caption{Dependence of (a) $E_\mathrm{a}$ and (b) $\ln \tilde{D}_0 + \ln N_\mathrm{SIA}$ on $\eta$ for various values of $N_\mathrm{SIA}$. The dotted line represents a linear fit to the data in each panel.}
    \label{fig:activation_energy_D0}
    \end{figure*}

Fig.~\ref{fig:self-climb_D} shows values of $D$ as functions of temperature for selected values of $N_\mathrm{SIA}$ (37 and 71). We note that only the range of relatively high temperatures is considered due to the large value of $E_\mathrm{m} = 2.359$ eV. The value of $D$ follows an Arrhenius relationship over the temperature range investigated. This trend is also observed for other values of $N_\mathrm{SIA}$. From these Arrhenius plots, we extract effective activation energies $E_\mathrm{a}$ for the overall self-climb process:
    \begin{equation}
    D = D_0 \exp{\left( - \frac{E_\mathrm{a}}{k_\mathrm{B} T} \right)} \text{,} \label{eq:26}
    \end{equation}
where $D_0$ is the pre-exponential factor. Fig.~\ref{fig:activation_energy_D0}~(a) shows that $E_\mathrm{a}$ depends linearly on $\eta$, as defined in Eq.~(\ref{eq:16}). Note that $\eta$ characterises the deviation of the ground state geometry from the ideal hexagon, as discussed above. We also find that the value of $D_\mathrm{0}$ can be expressed as a function of $N_\mathrm{SIA}$ and $\eta$ as
    \begin{equation}
    D_0 = \frac{A^{\eta}}{N_\mathrm{SIA}} D_0^* \, \text{.} \label{eq:27}
    \end{equation}
This yields
    \begin{equation}
    \ln \tilde{D}_0 + \ln N_\mathrm{SIA} = \ln \tilde{D}_0^* + \eta \ln A \, \text{.} \label{eq:28}
    \end{equation}
Here, $D_0^*$ and $A$ are the fitting parameters, and dimensionless notation convention is adopted, where the diffusion coefficients are defined as $\tilde{D}_0 = D_0/D_\mathrm{ref}$ and $\tilde{D}_0^* = D_0^*/D_\mathrm{ref}$, with $D_\mathrm{ref} = 1$ $\mathrm{nm}^2 \, \mathrm{s}^{-1}$. Fig.~\ref{fig:activation_energy_D0}~(b) shows that the value of $\ln \tilde{D}_0 + \ln N_\mathrm{SIA}$ exhibits a linear dependence on $\eta$, consistent with Eq.~(\ref{eq:27}). Previous atomistic modelling studies have reported that $E_\mathrm{a}$ is either independent of loop size \cite{Swinburne2016SciRep} or exhibits only a weak size dependence \cite{Okita2016ActaMater}. In contrast, our results suggest a notable size dependence, with $E_\mathrm{a}$ varying by more than 1 eV over the range of $N_\mathrm{SIA}$ explored in this study. Importantly, this variation does not follow a simple monotonic dependence on $N_\mathrm{SIA}$; rather, it is strongly correlated with the value of $\eta$. Together with the exponential dependence of $D_0$ on $\eta$, see Eq.~(\ref{eq:27}), these results show that configurational variations in the loop geometry play the central part in controlling the self-climb mobility of loops.

An earlier study for bcc Fe and W \cite{Swinburne2016SciRep} showed that the value of $E_\mathrm{a}$ was approximately twice the value of $E_\mathrm{m}$, which is the the migration barrier of SIA pipe-diffusion when $\Delta E_\mathrm{f} = 0$ (in Ref. \cite{Swinburne2016SciRep}, $E_a$ was interpreted as twice the vacancy migration energy in bulk $E_\mathrm{m}^\mathrm{V}$, since $E_\mathrm{m} = E_\mathrm{m}^\mathrm{V}$). In our study, we find that $E_\mathrm{m} = 2.359$ eV, resulting in $2E_\mathrm{m} = 4.718$ eV. This value is comparable to $E_\mathrm{a}$ of 4.847 eV found for $N_\mathrm{SIA} = 71$, indicating a good agreement with the findings of Ref. \cite{Swinburne2016SciRep}. Another study of bcc Fe \cite{Okita2016ActaMater} reported $E_\mathrm{a} = 1.67-1.98$ eV for loops with non-hexagonal shape, at the same time giving $E_\mathrm{m} = 1.02$ eV, again suggesting that $E_{a}$ is approximately twice the value of $E_\mathrm{m}$. Combining these results with the findings of our study suggests that if the loop ground-state configuration deviates from a perfect hexagon, the value of $E_\mathrm{a}$ can be reasonably well approximated as being twice that of $E_\mathrm{m}$ in bcc crystalline materials.

Our results also provide a new insight into the experimentally observed self-climb process. Fig.~\ref{fig:activation_energy_D0}~(a) suggests that the value of $E_\mathrm{a}$ varies by more than 1 eV depending on the variations of the ground state geometry, parameterised by $\eta$. Such a strong dependence implies that the self-climb processes observed experimentally are likely associated with loops with larger $\eta$ and hence lower $E_\mathrm{a}$. In contrast, loops with smaller $\eta$ and therefore higher $E_\mathrm{a}$ would not readily undergo self-climb under the same conditions. Instead, the occurrence of self-climb may be preceded by the absorption or emission of constituent SIAs from or to the bulk, thereby modifying $N_\mathrm{SIA}$ and consequently altering $\eta$. Through this configurational adjustment, $E_\mathrm{a}$ can be reduced to a level at which self-climb is initiated at a given temperature. This suggests that the inclusion of point defect absorption and emission to and from prismatic dislocation loops could further extend the quantitative modelling of the dislocation self-climb effects.


\section{Conclusions}

The main outcome of this study is that the thermodynamic properties and dynamics of dislocation loops are controlled by the morphological irregularity of their ground-state configurational geometry. In this work, we have established a quantitative link between the configuration of a loop and both its properties and dynamics. This conclusion has been demonstrated by performing three computational steps.

First, we have developed machine-learning models that predict the formation energy $E_\mathrm{f}$ of SIA configurations, spanning the idealised hexagon as well as geometrically complex arrangements of the SIA defects forming the loops. The predictive capacity of the model has been validated through comparison with atomistic simulations, demonstrating the high accuracy of the machine learning approach, with the typical error being within 1\% range.

Second, using these models, we have evaluated the density of microstates $g(E_\mathrm{f})$ for various SIA configurations and derived an analytical model for $\ln g(E_\mathrm{f})$. We have discovered that a critical parameter $\eta$, characterising the morphological irregularity of the ground-state configuration and its deviation from a perfect hexagon, critically affects the function $\ln g(E_\mathrm{f})$. Furthermore, the analytical model offers a powerful framework for investigating the thermodynamic properties of geometrically complex loops: once the ground-state information is established, key thermodynamic quantities can be derived using the well-established framework of statistical mechanics. The free energy $F$, mean energy $\langle E_\mathrm{f} \rangle$, and entropy $S$ have been evaluated for the various values of $N_\mathrm{SIA}$. To prove that these quantities are controlled by the value of $\eta$, we have derived analytical expressions for the thermodynamic quantities as functions of this parameter.

Third, we have simulated the self-climb process of a dislocation loop using the machine-learning models. We have shown that the activation energy $E_\mathrm{a}$ and the pre-exponential factor $D_0$ exhibit clearly identifiable variation as functions of $\eta$, proving that morphological variations of the ground-state geometry play the key part in loop dynamics.

We note that loop thermodynamic properties and dynamics were investigated using fundamentally different modelling techniques --- the Wang--Landau sampling and atomistic kMC simulations, respectively. Nevertheless, both approaches produced a consistent conclusion: {\it both} the loop thermodynamic properties {\it and} dynamics can be quantitatively represented as functions of $\eta$, which characterises the ground-state loop geometry. This remarkable agreement provides compelling proof of both the validity of our central conclusion and the robustness of the modelling approaches adopted in this study. Taken together, these results show that $\eta$ transcends its role as a mere parameter and instead represents a physically meaningful quantity controlling both the thermodynamics and dynamics of prismatic dislocation loops.


\section*{Conflicts of interest}

There are no conflicts to declare.

\section*{Data availability}

Data will be made available on request. The code used in the current study will be made available in an open-source repository.

\section*{Acknowledgements}
This work has been carried out within the framework of the EUROfusion Consortium, funded by the European Union via the Euratom Research and Training Programme (Grant Agreement No. 101052200—EUROfusion) and was partially supported by the Broader Approach Phase II agreement under the PA of IFERC2-T2PA02. This work was also funded by the EPSRC Energy Programme (Grant No. EP/W006839/1).  Views and opinions expressed are, however, those of the authors only and do not necessarily reflect those of the European Union or the European Commission. Neither the European Union nor the European Commission can be held responsible for them.
The authors acknowledge the use of the Cambridge Service for Data Driven Discovery (CSD3) and associated support services provided by the University of Cambridge Research Computing Services (https://www.csd3.cam.ac.uk) in the completion of this work. To obtain further information on the data and models underlying this paper, please contact PublicationsManager@ukaea.uk.

\section*{Author contributions} 
S.H. conceived the study, developed the methodology, performed the simulations, conducted formal analysis, curated the data, and wrote the original draft. S.L.D. contributed to conceptual development and interpretation of the results. M.B. contributed to data analysis and interpretation. All authors reviewed and edited the manuscript and approved the final version.

\bibliography{references}

\end{document}


\title{Supplementary Materials for:\\
Thermodynamics and dynamics of non-compact prismatic dislocation loops simulated using a machine-learning model}

\author{Sho Hayakawa}
\author{Sergei L. Dudarev}
\author{Max Boleininger}
\maketitle

\section*{I. Proof of the uniqueness of $\boldsymbol{n}_i$ defined by Eq. (8)}
We first derive the general representation of the position of the $i$th atomic string $\boldsymbol{r}_i$. Consider the following two representations:
    \begin{align}
    \boldsymbol{r}_i &= \sum_k m_{i,k} \boldsymbol{e}_k = m_{i,1} \boldsymbol{e}_1 + m_{i,2} \boldsymbol{e}_2 + m_{i,3} \boldsymbol{e}_3 \text{,} \\
    \boldsymbol{r}_i &= \sum_k m'_{i,k} \boldsymbol{e}_k = m'_{i,1} \boldsymbol{e}_1 + m'_{i,2} \boldsymbol{e}_2 + m'_{i,3} \boldsymbol{e}_3 \text{,}
    \end{align}
where $m_{i,k}$ and $m'_{i,k}$ are integers. Note that $m_{i,k}$ does not necessarily equal $m'_{i,k}$ due to the redundancy of the basis vectors $\boldsymbol{e}_k$.
Since $\sum_k m_{i,k} \boldsymbol{e}_k = \sum_k m'_{i,k} \boldsymbol{e}_k$, we obtain
    \begin{equation}
    (m_{i,1} - m'_{i,1}) \boldsymbol{e}_1 + (m_{i,2} - m'_{i,2}) \boldsymbol{e}_2 + (m_{i,3} - m'_{i,3}) \boldsymbol{e}_3 = 0 \text{.} \label{eq:s3}
    \end{equation}
Using $\boldsymbol{e}_3 = -\boldsymbol{e}_1 - \boldsymbol{e}_2$, Eq. (\ref{eq:s3}) becomes
    \begin{equation}
    [(m_{i,1} - m'_{i,1}) - (m_{i,3} - m'_{i,3})] \boldsymbol{e}_1 + [(m_{i,2} - m'_{i,2}) - (m_{i,3} - m'_{i,3})] \boldsymbol{e}_2 = 0 \text{.}
    \end{equation}
Because $\boldsymbol{e}_1$ and $\boldsymbol{e}_2$ are linearly independent, it follows that
    \begin{align}
    (m_{i,1} - m'_{i,1}) - (m_{i,3} - m'_{i,3}) &= 0 \text{,} \\
    (m_{i,2} - m'_{i,2}) - (m_{i,3} - m'_{i,3}) &= 0 \text{.}
    \end{align}
Therefore,    
    \begin{equation}
    m_{i,1} - m'_{i,1} = m_{i,2} - m'_{i,2} = m_{i,3} - m'_{i,3} \text{.}
    \end{equation}
Consequently, the general representation of $\boldsymbol{r}_i$ can be written as
    \begin{equation}
    \boldsymbol{r}_i = \sum_k (m_{i,k} + t) \boldsymbol{e}_k = (m_{i,1} + t) \boldsymbol{e}_1 + (m_{i,2} + t) \boldsymbol{e}_2 + (m_{i,3} + t) \boldsymbol{e}_3 \text{.}
    \end{equation}
where $t$ is an arbitrary integer.

According to Eq. (8) in the main text, to obtain a unique representation, we determine the value of $t$ that minimises
    \begin{equation}
    f(t) = | m_{i,1} + t | + | m_{i,2} + t | + | m_{i,3} + t | \text{.}
    \end{equation}
We then define
    \begin{equation}
    \boldsymbol{n}_i = (m_{i,1} + t_\mathrm{min}, m_{i,2} + t_\mathrm{min}, m_{i,3} + t_\mathrm{min}) \text{,}
    \end{equation}
where $t_\mathrm{min}$ is the value of $t$ that minimises $f(t)$.
Without loss of generality, assume $m_{i,1} \le m_{i,2} \le m_{i,3}$. Then, 
    \begin{align}
    f(-m_{i,2}) &= | m_{i,1} - m_{i,2} | + | m_{i,3} - m_{i,2} |  \nonumber \\
    &= (m_{i,2} - m_{i,1}) + (m_{i,3} - m_{i,2}) \nonumber \\
    &= m_{i,3} - m_{i,1} \text{.}
    \end{align}
Moreover, by the triangle inequality, 
    \begin{align}
    | m_{i,1} + t | + | m_{i,3} + t | &\ge | (m_{3,1} + t) - (m_{i,1} + t)| \nonumber \\
    &= m_{i,3} - m_{i,1} \text{.}
    \end{align}
Thus, 
    \begin{align}
    f(t) &= | m_{i,1} + t | + | m_{i,2} + t | + | m_{i,3} + t | \nonumber \\
    &\ge | m_{i,2} + t | + (m_{i,3} - m_{i,1}) \nonumber \\
    &= | m_{i,2} + t | + f(-m_{i,2}) \text{.}
    \end{align}
Hence, $f(t) \ge\ f(-m_{i,2})$, and equality holds if and only if $t = -m_{i,2}$. Therefore, $t = -m_{i,2}$ uniquely minimises $f(t)$.

\section*{II. Influence of machine-learning architecture on model performance}

\begin{figure*}[htbp]
    \centering
    \includegraphics[width=1.0\textwidth]{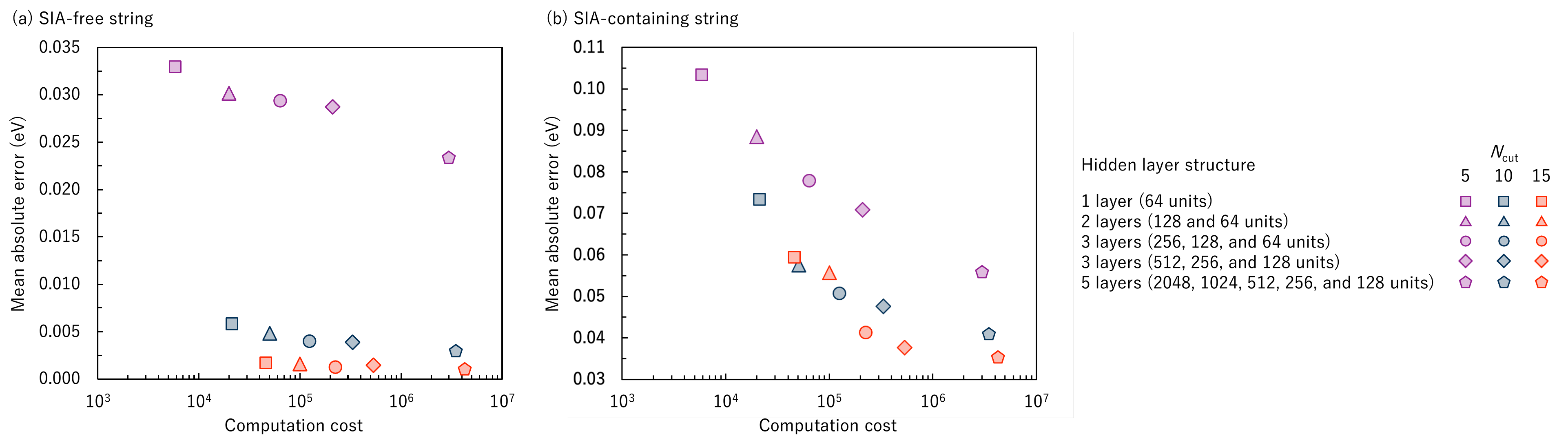}
    \caption{Dependence of MAE as a function of computation cost for different model architectures: (a) SIA-free string and (b) SIA-containing string.}
    \label{fig:computation_cost}
    \end{figure*}

Figs. \ref{fig:computation_cost} (a) and (b) show the dependence of the mean absolute error (MAE) on the structure of the hidden layers for the SIA-free and SIA-containing string models, respectively.
Here, the MAE values are plotted as a function of the computation cost required for a model to generate an output ($E^{i}_{[111]}$) from a given input (local SIA occupancy pattern).
This cost originates from the dense matrix-vector multiplications in each fully connected layer, i.e., the products of the weight matrices with the input and hidden activation vectors.
It is estimated as
    \begin{equation}
    \mathrm{Cost} = N^\mathrm{in}_\mathrm{unit} N^1_\mathrm{unit} + \sum^{N_\mathrm{hidden}-1}_k N^{k}_\mathrm{unit} N^{k+1}_\mathrm{unit} + N^{N_\mathrm{hidden}}_\mathrm{unit} N^\mathrm{out}_\mathrm{unit} \text{,}
    \end{equation}
where $N^\mathrm{in}_\mathrm{unit}$, $N^\mathrm{out}_\mathrm{unit}$, and $N^k_\mathrm{unit}$ denote the numbers of units in the input layer, output layer, and $k$th hidden layer, respectively, and $N_\mathrm{hidden}$ is the number of hidden layers.
The values of $N^\mathrm{in}_\mathrm{unit}$ are 91, 331, and 721 for $n_\mathrm{cut}$ = 5, 10, and 15, respectively, and $N^\mathrm{out}_\mathrm{unit} = 1$ for all models.
As shown in the figures, a deeper and wider hidden-layer structure leads to higher predictive accuracy for both the SIA-free and SIA-containing strings at $n_\mathrm{cut}$ = 5, arising from the greater flexibility of the model, albeit at the expense of increased computation cost. 
In contrast, the influence of the network depth and width becomes small for the SIA-free string case at $n_\mathrm{cut}$ = 10 and 15.
This suggests that even a relatively simple model is sufficient to capture the inherent relationship between $E^{i}_{[111]}$ and the local SIA-occupancy pattern for an SIA-free string when $n_\mathrm{cut}$ is sufficiently large, possibly owing to the relatively narrow energy range of $E^{i}_{[111]}$ (Figs. 3 (a-1)--(a-3) in the main text).
For the SIA-containing string case, while noticeable improvements in accuracy are observed when increasing the hidden-layer depth and width, the MAE appears to start saturating at a computation cost of approximately $10^6$ in some cases.
For instance, the MAE difference between the three-layer model with 256, 128, and 64 units (circle symbols) and the five-layer model (pentagon symbols) at $n_\mathrm{cut}$ = 15 is only 0.006 eV, whereas the computation cost differs by a factor of 20.
Thus, selecting an optimal network structure requires balancing accuracy and computation cost according to the desired precision for the problem of interest.
For the analysis in the main text, we adopt the three-layer model with 256, 128, and 64 units as it offers a good balance between accuracy and computational efficiency.

\section*{III. Merging results from neighbouring energy ranges in multi-range Wang--Landau sampling}

The results from neighbouring ranges are merged using the following procedure.
First, we compute the average difference in $\mathrm{ln} \,  g(E_\mathrm{f})$ across the overlapping energy bins between the $i$th and $({i - 1)}$th ranges:
    \begin{equation}
    \Delta_i = \frac{1}{N_{\mathrm{overlap},i}} \sum_{E_\mathrm{f} \in \mathrm{overlap}} \{\mathrm{ln} \, g_i(E_\mathrm{f}) - \mathrm{ln} \, g_{i-1}(E_\mathrm{f})\} \text{,}
    \end{equation}
where $g_i(E_\mathrm{f})$ denotes the $g(E_\mathrm{f})$ obtained in the $i$th range and $N_{\mathrm{overlap},i}$ is the number of the overlapping bins between the ${i}$th and $({i - 1)}$th ranges.
We then shift $\mathrm{ln} \, g_i(E_\mathrm{f})$ by subtracting $\Delta_i$:
    \begin{equation}
    \mathrm{ln} \, g_i(E_\mathrm{f}) \leftarrow \mathrm{ln} \, g_i(E_\mathrm{f}) - \Delta_i
    \end{equation}
Note that subtracting a constant from $\mathrm{ln} \, g(E_\mathrm{f})$ does not alter the physical information contained in $g(E_\mathrm{f})$ because the Wang--Landau sampling determines only the relative values of $g(E_\mathrm{f})$.
Finally, the merged values of $g(E_\mathrm{f})$ in the overlap region are obtained using the weighted average:
    \begin{equation}
    \mathrm{ln} \, g(E_\mathrm{f}) = w(E_\mathrm{f}) \ \mathrm{ln} \, g_{i-1}(E_\mathrm{f}) + \{1-w(E_\mathrm{f})\} \, \mathrm{ln} \, g_i(E_\mathrm{f}) \text{,}
    \end{equation}
where $w(E_\mathrm{f})$ is a weight function that decreases linearly from 1 at the low-energy edge of the overlap to 0 at the high-energy edge.

After merging the neighbouring ranges, we shift the entire $g_i(E_\mathrm{f})$ so that it represents absolute values.
To achieve this, we count the number of unique configurations that fall into the minimum energy bin in Range 1 during the sampling ($N_\mathrm{unique}^\mathrm{min}$) while accounting for translational invariance of configurations in terms of $E_\mathrm{f}$.
The values of $g_i(E_\mathrm{f})$ are then rescaled as:
    \begin{equation}
    \mathrm{ln} \, g(E_\mathrm{f}) \leftarrow \mathrm{ln} \, g(E_\mathrm{f}) - \mathrm{ln} \, g(E_\mathrm{f}^\mathrm{min}) + \mathrm{ln} \, N_\mathrm{unique}^\mathrm{min} \text{,}
    \end{equation}
where $E_\mathrm{f}^\mathrm{min}$ denotes the minimum $E_\mathrm{f}$ in Range 1.

{
\setlength{\tabcolsep}{10pt}

\begin{longtable}{ccc}
\caption{Energy ranges of $E_\mathrm{f}$ employed in the multi-range Wang--Landau sampling.}
\label{tab:Ef_ranges}\\

\toprule
Range ID & Min.\ $E_\mathrm{f}$ (eV) & Max.\ $E_\mathrm{f}$ (eV) \\
\midrule
\endfirsthead

\toprule
Range ID & Min.\ $E_\mathrm{f}$ (eV) & Max.\ $E_\mathrm{f}$ (eV) \\
\midrule
\endhead

\bottomrule
\endlastfoot

\multicolumn{3}{l}{$N_\mathrm{SIA}=37$}\\
  1 & 118 & 125 \\
  2 & 120 & 140 \\
  3 & 130 & 210 \\ 
  4 & 200 & 310 \\ 
  5 & 300 & 330 \\
\addlinespace

\multicolumn{3}{l}{$N_\mathrm{SIA}=50$}\\
  1 & 145 & 160 \\
  2 & 150 & 210 \\
  3 & 200 & 310 \\ 
  4 & 300 & 410 \\ 
  5 & 400 & 450 \\
\addlinespace

\multicolumn{3}{l}{$N_\mathrm{SIA}=71$}\\
  1 & 183 & 210 \\
  2 & 200 & 310 \\
  3 & 300 & 410 \\ 
  4 & 400 & 510 \\ 
  5 & 500 & 610 \\ 
  6 & 600 & 630 \\
\addlinespace

\multicolumn{3}{l}{$N_\mathrm{SIA}=91$}\\
  1 & 213 & 225 \\
  2 & 220 & 240 \\
  3 & 230 & 310 \\ 
  4 & 300 & 410 \\ 
  5 & 400 & 510 \\ 
  6 & 500 & 610 \\ 
  7 & 600 & 710 \\ 
  8 & 700 & 810 \\
\addlinespace

\multicolumn{3}{l}{$N_\mathrm{SIA}=110$}\\
  1 & 241 & 260 \\
  2 & 250 & 310 \\
  3 & 300 & 410 \\ 
  4 & 400 & 510 \\ 
  5 & 500 & 610 \\ 
  6 & 600 & 710 \\ 
  7 & 700 & 810 \\ 
  8 & 800 & 910 \\ 
  9 & 900 & 1010 \\
\addlinespace

\multicolumn{3}{l}{$N_\mathrm{SIA}=130$}\\
  1 & 268 & 280 \\
  2 & 275 & 310 \\
  3 & 300 & 410 \\ 
  4 & 400 & 510 \\ 
  5 & 500 & 610 \\ 
  6 & 600 & 710 \\ 
  7 & 700 & 810 \\ 
  8 & 800 & 910 \\ 
  9 & 900 & 1010 \\ 
  10 & 1000 & 1110 \\ 
  11 & 1100 & 1160 \\
\addlinespace

\multicolumn{3}{l}{$N_\mathrm{SIA}=150$}\\
  1 & 292 & 300 \\
  2 & 295 & 310 \\
  3 & 300 & 410 \\ 
  4 & 400 & 510 \\ 
  5 & 500 & 610 \\ 
  6 & 600 & 710 \\ 
  7 & 700 & 810 \\ 
  8 & 800 & 910 \\ 
  9 & 900 & 1010 \\ 
  10 & 1000 & 1110 \\ 
  11 & 1100 & 1210 \\ 
  12 & 1200 & 1310 \\ 
  13 & 1300 & 1360 \\
\addlinespace

\multicolumn{3}{l}{$N_\mathrm{SIA}=169$}\\
  1 & 315 & 323 \\
  2 & 318 & 350 \\
  3 & 340 & 410 \\ 
  4 & 400 & 510 \\ 
  5 & 500 & 610 \\ 
  6 & 600 & 710 \\ 
  7 & 700 & 810 \\ 
  8 & 800 & 910 \\ 
  9 & 900 & 1010 \\ 
  10 & 1000 & 1110 \\ 
  11 & 1100 & 1210 \\ 
  12 & 1200 & 1310 \\ 
  13 & 1300 & 1410 \\ 
  14 & 1400 & 1510 \\

\end{longtable}

}

\FloatBarrier

\clearpage

\section*{IV. Influence of $L$ on $p, q$ and $\eta$}

\begin{figure}[htbp]
    \centering
    \includegraphics[width=0.9\linewidth]{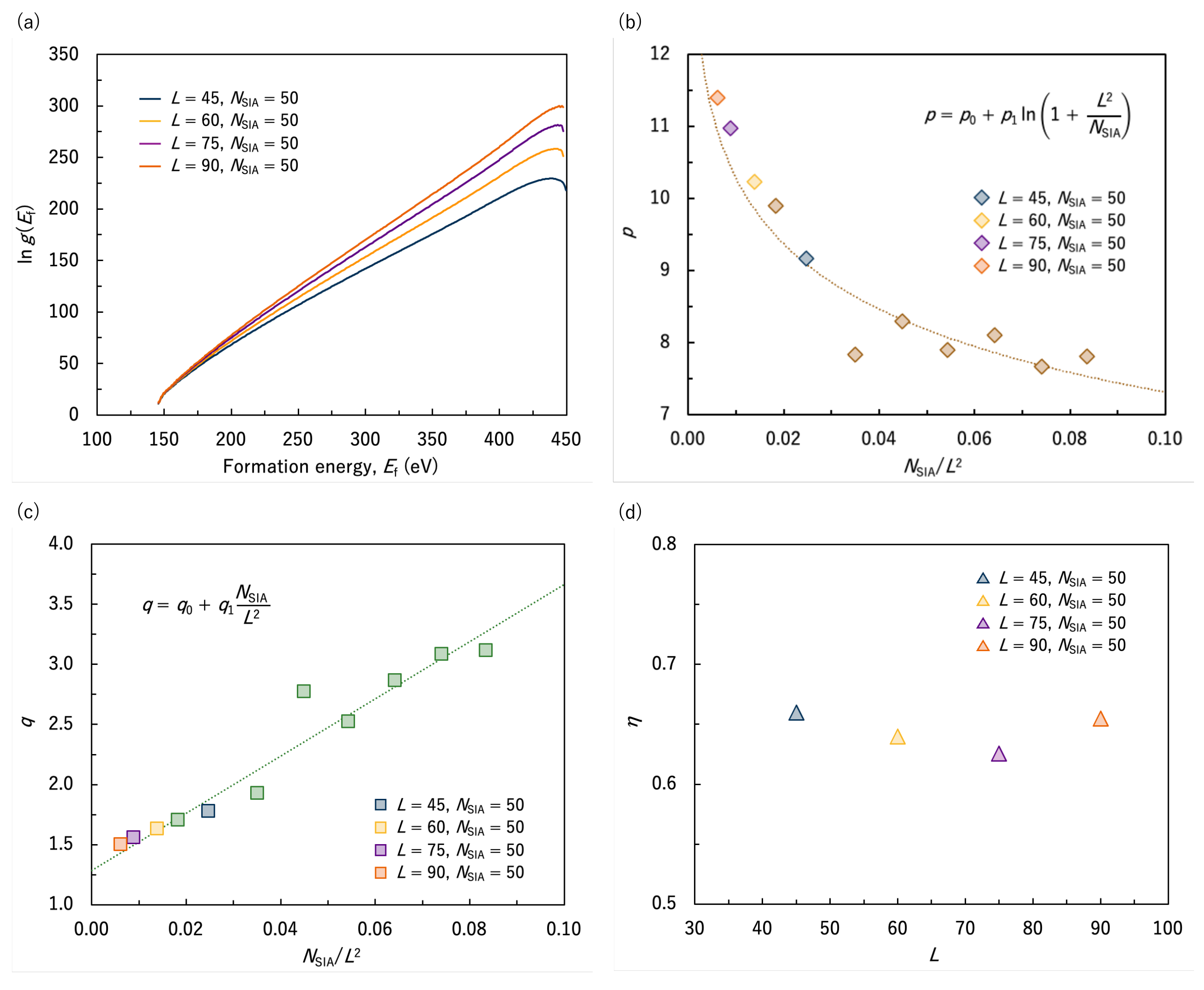}
    \caption{(a) Values of $g (E_\mathrm{f})$ as a function of $E_\mathrm{f}$ for various values of $L$ with $N_\mathrm{SIA} = 50$ fixed. (b) Fitted values of $p$ as a function of $N_\mathrm{SIA}/L^2$. (c) Fitted values of $q$ as a function of $N_\mathrm{SIA}/L^2$. (d) Fitted values of $\eta$ as a function of $L$.}
    \label{fig:L_dependence}
    \end{figure}

Fig. \ref{fig:L_dependence} (a) shows $g(E_\mathrm{f})$ as a function of $E_\mathrm{f}$ for various values of $L$ with $N_\mathrm{SIA} = 50$ fixed. Following Eq. (13) in the main text, we fit the parameters $p, q$, and $\eta$ to these data and evaluate their dependences on $L$, as presented in Figs. \ref{fig:L_dependence} (b)--(d). Note, in Figs. \ref{fig:L_dependence} (b) and (c), the obtained values of $p$ and $q$ are overlaid with the data points and fitted curves in Fig. \hl{9} (c) and (d) of the main text, respectively. The values of $p$ and $q$ at different $L$ are found to be consistent with the dependencies described by Eqs. (14) and (15) in the main text, supporting the validity of these relationships.
In contrast, we find that $\eta$ shows no clear dependence on $L$, as demonstrated in Fig. \ref{fig:L_dependence} (d).
This is likely because, as discussed in the main text, $\eta$ corresponds to the morphological irregularity of the ground state configuration, in which all SIAs are continuously connected in a relatively compact geometry.
Consequently, $\eta$ exhibits no dependence on the system size.

\section*{V. Definitions of $P$ and $R_\mathrm{c}$}

\begin{figure}[htbp]
    \centering
    \includegraphics[width=0.4\linewidth]{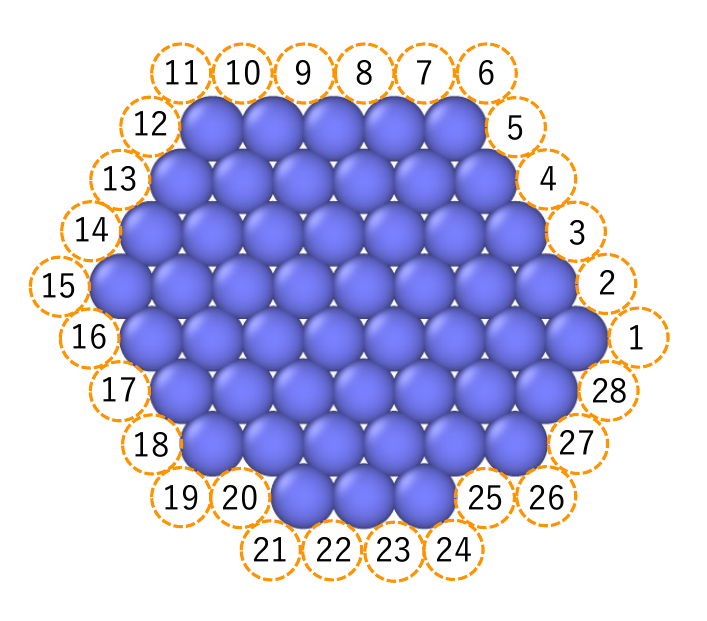}
    \caption{Ground state configuration for $N_\mathrm{SIA} = 50$. The dashed orange circles represent the SIA-free strings located along the edge of the configuration, whose total number defines the perimeter $P$.}
    \label{fig:P_definition}
    \end{figure}

The value of $P$ is defined as the number of SIA-free strings located around the edge of the ground-state configuration, representing its perimeter.
As an example, Fig. \ref{fig:P_definition} illustrates the ground-state configuration for $N_\mathrm{SIA}$ = 50, where 28 SIA-free strings surround the configuration. Accordingly, $P = 28$ in this case.

We define the characteristic radius of the ground state configuration $R_\mathrm{c}$, by requiring that it satisfies the centred hexagonal number relation
    \begin{equation}
    3 R_\mathrm{c}^2 - 3 R_\mathrm{c} + 1 = N_\mathrm{SIA} \text{.} \label{eq:s5}
    \end{equation}
Solving Eq. (\ref{eq:s5}) for $R_\mathrm{c}$ gives
    \begin{equation}
    R_\mathrm{c} = \frac{3 + \sqrt{12 N_\mathrm{SIA} - 3}}{6} \text{.} \label{eq:s6}
    \end{equation}
When the configuration forms a perfect hexagon, i.e., when $N_\mathrm{SIA}$ is a centred hexagonal number, $R_\mathrm{c}$ is a positive integer and corresponds to the circumradius of the hexagon.
For non-magic numbers of $N_\mathrm{SIA}$, $R_\mathrm{c}$ is non-integer; nevertheless, it can be interpreted as the circumradius of a hypothetical perfect hexagon with the same $N_\mathrm{SIA}$.

Under these definitions, the value of $P/R_\mathrm{c}$ serves as a size-independent measure of the perimeter.
For a perfect hexagonal configuration, one obtains $P/R_\mathrm{c} = 6$.
Consequently, the quantity $P/R_\mathrm{c} - 6$ provides a direct measure of the morphological deviation from the ideal hexagonal geometry.

\revA{\section*{VI. Derivation of the functional form of the dependence of $\ln g (E_\mathrm{f})$ on $\xi$ and assessment of the fitting quality}}

\revA{As discussed in the main text, the functional form adopted in Eq. (\hl{12}) is motivated by the observed dependence of $\frac{d}{d\xi} \ln g(E_\mathrm{f})$ on $\xi$: the value varies only weakly over the intermediate $\xi$ range, whereas it rapidly increases as $\xi$ approaches zero. We therefore assume the following functional form in the low to intermediate $\xi$ range:
    \begin{equation}
    \frac{d}{d\xi} \ln g(E_\mathrm{f}) = A_0 + A_1 \xi^{\eta^{\prime}} \text{,}
    \end{equation}
where $A_0$, $A_1$, and $\eta^{\prime}$ are fitting parameters, with $A_0, A_1 > 0$ and $-1 < \eta^{\prime} < 0$. Introducing the parameters, $p$, $q$, $r$, and $\eta$, we reparameterise the above expression by setting
    \begin{align}
    \eta^{\prime} &= \eta - 1 \text{,} \\
    A_0 &= p - r\eta \text{} \\
    A_1 &= qr^{\eta} \text{.}
    \end{align}
This reparameterisation is selected empirically after exploring several alternatives, with the aim of obtaining simple and systematic dependences of $p$ and $q$ on $N_\mathrm{SIA}/L^2$, and of $\eta$ on $P/R_\mathrm{C}-6$. This gives
    \begin{equation}
    \frac{d}{d\xi} \ln g(E_\mathrm{f}) = p - r\eta + \frac{qr^{\eta}}{\xi^{1-\eta}} \text{,}
    \end{equation}
where $p > r\eta$, $q > 0$, $r > 0$, and $0 < \eta < 1$. This expression is identical to Eq.~(\hl{12}) in the main text.}

\revA{As described in the main text, integrating Eq. (\hl{12}) with respect to $\xi$ yields the analytical expression for $\ln g(E_\mathrm{f})$ given in Eq. (\hl{13}). To quantitatively assess the quality of this analytical representation, we calculate the MAE of the predicted $\ln g (E_\mathrm{f})$ relative to the Wang--Landau data over the low- to intermediate $\xi$ range considered in the present analysis. The resulting MAEs for each $N_\mathrm{SIA}$ are summarised in Table \ref{tab:fit_quality_eq13}, demonstrating the high predictive accuracy of the analytical representation.}

{
\setlength{\tabcolsep}{20pt}

\begin{table}[htbp]
\centering
\caption{\revA{MAEs of the analytical representation of $\ln g (E_\mathrm{f})$ given by Eq. (\hl{13}) for each $N_\mathrm{SIA}$, evaluated relative to the Wang--Landau data over the low to intermediate $\xi$ range considered in the present analysis.}}
\label{tab:fit_quality_eq13}

\begin{tabular}{cc}
\toprule
\revA{$N_\mathrm{SIA}$} & \revA{MAE} \\
\midrule
\revA{37}    & \revA{0.350} \\
\revA{50}    & \revA{0.327} \\
\revA{71}    & \revA{0.308} \\
\revA{91}    & \revA{0.644} \\
\revA{110}    & \revA{0.623} \\
\revA{130}    & \revA{0.886} \\
\revA{150}    & \revA{0.743} \\
\revA{169}    & \revA{0.922} \\
\bottomrule
\end{tabular}

\end{table}
}

    
\revA{\section*{VII. Validation for the functional forms chosen for $p$, $q$, and $\eta$}}

\revA{The functional forms for the parameters $p$, $q$, and $\eta$ in Eqs. (\hl{14}), (\hl{15}), and (\hl{16}) of the main text, respectively, are chosen phenomenologically. To assess the validity of these choices, we perform symbolic regression for each parameter using PySR \cite{cranmer2023pysr} and compare the resulting functional forms with those adopted in the main text. The search space includes the binary operators $+,-,\times,$ and $/$, and the unary operators $\exp$ and $\ln$.}

\begin{figure}[htbp]
    \centering
    \includegraphics[width=1.0\linewidth]{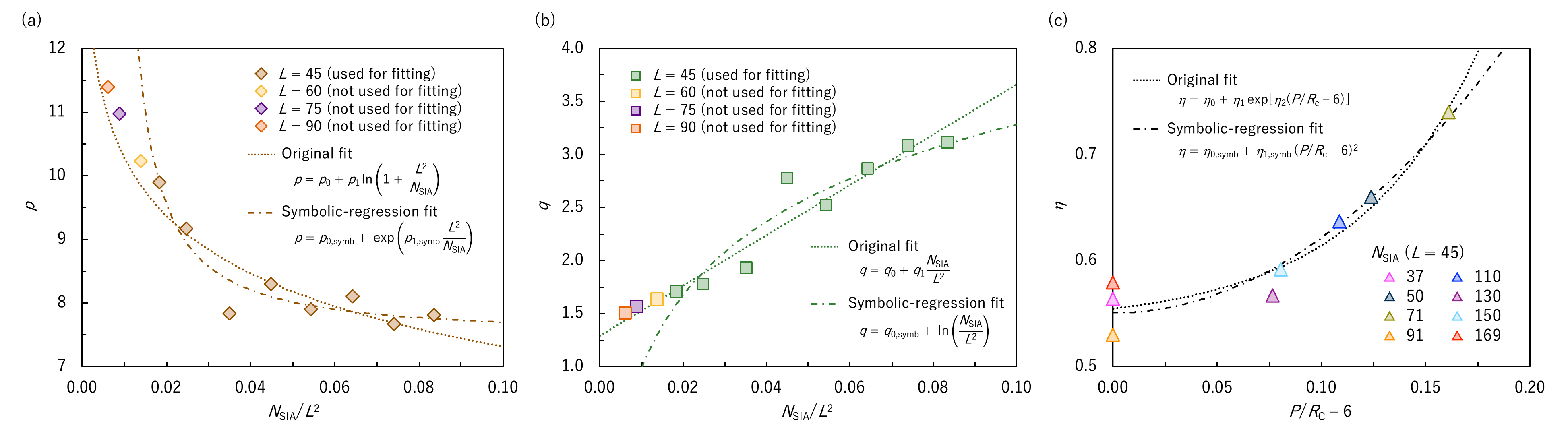}
    \caption{\revA{Comparison of the original and symbolic regression fitting expressions for (a) $p$, (b) $q$, and (c) $\eta$. The data for $L =$ 45, which are used for fitting, are shown in each panel. For comparison, the $L =$ 60, 75, and 90 data, which are not included in the fitting, are also shown in (a) and (b).}}
    \label{fig:symbolic_regression}
    \end{figure}

\revA{The dotted-dashed curves in Fig. \ref{fig:symbolic_regression} (a)--(c) show the fits obtained using the functional forms identified by symbolic regression for $p$, $q$, and $\eta$, respectively. To ensure a direct comparison under the same fitting conditions, the symbolic-regression analysis is performed using only the data for $L =$ 45, as in the original fitting. The functional forms selected on the basis of the balance between fitting accuracy and functional complexity are
    \begin{align}
    p &= p_\mathrm{0,symb} + \exp{\left(p_\mathrm{1,symb}\frac{L^2}{N_\mathrm{SIA}}\right)} \text{,} \\
    q &= q_\mathrm{0,symb} + \ln{\left(\frac{N_\mathrm{SIA}}{L^2} \right)} \text{,} \\
    \eta &= \eta_\mathrm{0,symb} + \eta_\mathrm{1,symb} \left( \frac{P}{R_\mathrm{C}} - 6 \right)^2 \text{,}
    \end{align}
where $p_\mathrm{0,symb}$, $p_\mathrm{1,symb}$, $q_\mathrm{0,symb}$, $\eta_\mathrm{0,symb}$, and $\eta_\mathrm{1,symb}$ are fitted as 6.437, 0.023, 5.586, 0.550, and 7.098, respectively.}

\revA{For $p$, symbolic regression identifies a functional form different from that adopted in Eq. (\hl{14}), while both expressions contain two fitted parameters. For the $L =$ 45 data used in the fitting, the mean absolute errors (MAEs) are 0.252 and 0.182 for the original and symbolic-regression fits, respectively, indicating a modest improvement in fitting accuracy with the symbolic-regression form. However, comparison with the  $L =$ 60, 75, and 90 data, none of which are included in the fitting, shows that the original functional form retains good predictive accuracy, whereas the symbolic-regression form exhibits substantially larger deviations (see the yellow, purple, and orange points in the figure). This comparison suggests that, despite its slightly larger fitting error for $L =$ 45, the original functional form provides better transferability across different values of $L$, supporting the validity of the adopted form.}

\revA{For $q$, symbolic regression likewise identifies a functional form different from that adopted in Eq. (\hl{15}). The MAEs for the $L =$ 45 data are 0.128 and 0.140 for the original and symbolic-regression form, respectively, indicating slightly better fitting accuracy for the original form. The symbolic-regression expression contains only one fitted parameter, compared with two in the original expression, and might therefore be considered somewhat simpler. However, comparison with the $L =$ 60, 75, and 90 outside the fitting set shows that the symbolic-regression form has poorer predictive performance, whereas the original linear relation retains excellent agreement. This result indicates that the linear dependence adopted in Eq. (\hl{15}) provides better transferability across different values of $L$, as also observed for $p$, thereby supporting the validity of the adopted forms.}

\revA{For $\eta$, symbolic regression again identifies a functional form different from that adopted in Eq. (\hl{16}). However, the resulting fitting curves are qualitatively very similar for the original and symbolic-regression expressions. For the $L =$ 45 data, the MAEs are 0.014 and 0.013 for the original and symbolic-expression forms, respectively, indicating essentially comparable fitting accuracy. The original expression contains three fitted parameters, whereas the symbolic-regression expression contains only two and is therefore slightly simpler. Nevertheless, the negligible difference in accuracy suggests that the original form remains fully adequate. Importantly, the detailed choice of functional form does not significantly affect the discussion or conclusions of the present study, provided that the adopted expression reproduces the data with sufficient accuracy.}

\FloatBarrier


\revA{\section*{VIII. Sensitivity analysis of the fits for $p$, $q$, and $\eta$}}

\begin{figure}[htbp]
    \centering
    \includegraphics[width=1.0\linewidth]{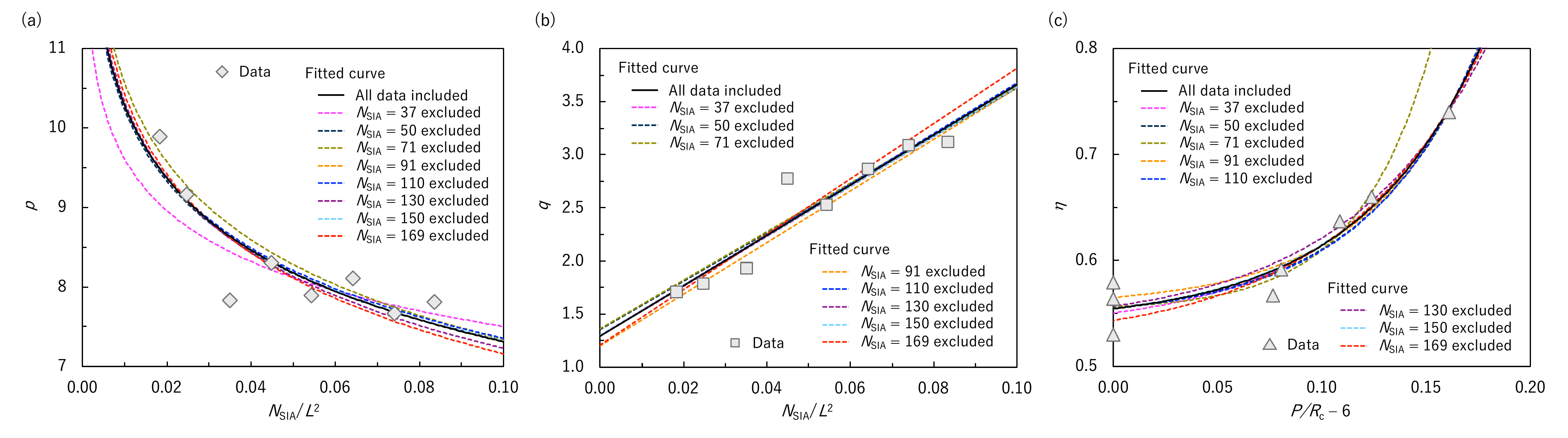}
    \caption{\revA{Leave-one-out (LOO) sensitivity analysis of the fitted functions for (a) $p$, (b) $q$, and (c) $\eta$. Each coloured dashed curve represents the fit obtained by excluding the indicated data point, while the black solid curve represents the fit obtained using all data points. The data points used for the fitting are also shown in each panel.}}
    \label{fig:sensitivity_analysis}
    \end{figure}

\revA{To examine the sensitivity of the fitted functions for $p$, $q$, and $\eta$ (Eqs. (\hl{14})--(\hl{16})) to individual data points, we perform leave-one-out (LOO) analysis, in which each data point is excluded in turn and the function is refitted to the remaining data. This analysis quantifies the extent to which the fitted function depends on any single data point. Figs. \ref{fig:sensitivity_analysis} (a)--(c) show the resulting fitted functions obtained by excluding each data point individually. We here define the LOO error $e_i^\mathrm{LOO}$ for the $i$th data point as
    \begin{equation}
    e_i^\mathrm{LOO} = \left| y_i - f_i (x_i) \right| \text{,}
    \end{equation}
where $x_i$ and $y_i$ denote the input variable and corresponding observed value of the $i$th data point, respectively, and $f_i(x_i)$ is the value predicted at $x_i$ by the function fitted after excluding that data point. The LOO error therefore quantifies how accurately an individual data point can be predicted from the remaining data when that point is excluded from the fitting. The mean absolute LOO error is then defined as
    \begin{equation}
    \mathrm{MAE}_\mathrm{LOO} = \frac{1}{N} \sum_{i=1}^{N} e_i^\mathrm{LOO} \text{,}
    \end{equation}
where $N = 8$ is the number of data points. The resulting LOO statistics, including the mean and maximum LOO errors, are summarised in Table \ref{tab:fit_sensitivity}, together with the MAEs obtained from the full-data fits.}

\revA{As shown in Fig. \ref{fig:sensitivity_analysis} (b), the fitted function for $q$ exhibits little sensitivity to the exclusion of individual data points. This is also reflected in the comparative MAE values obtained from the full-data and LOO analyses, as summarised in Table \ref{tab:fit_sensitivity}. These results indicate that the fitted function for $q$ is robust with respect to the individual data points and retains good predictive capability.}

\revA{In contrast, the fitted functions for $p$ and $\eta$ show sensitivity to the exclusion of certain data points, as seen in Fig. \ref{fig:sensitivity_analysis} (a) and (c), respectively. For $p$, a relatively large deviation is observed when the $N_\mathrm{SIA} = 37$ point is excluded, while for $\eta$, the largest deviation occurs when the $N_\mathrm{SIA} = 71$ point is excluded. Importantly, these points lie at the boundaries of the sampled range: $N_\mathrm{SIA} = 37$ corresponds to the smallest $N_\mathrm{SIA} / L^2$ value, 0.018, whereas $N_\mathrm{SIA} = 71$ corresponds to the largest $P/R_\mathrm{C}-6$ value, 0.161. Excluding such boundary points reduces the constraint on the fitted function near the corresponding edge of the sampled range and effectively creates an extrapolative region, where large deviations are therefore expected. These large deviations therefore primarily reflect extrapolation beyond the range constrained by the remaining data, rather than a deterioration in the fitting quality within the sampled range. Indeed, except for these boundary cases, the excluded data points are generally well predicted from the remaining data. We therefore consider the fitted functions for $p$ and $\eta$ to exhibit sufficient robustness overall, while showing a natural sensitivity associated with these boundary cases.}

{
\setlength{\tabcolsep}{10pt}

\begin{table}[htbp]
\centering
\caption{\revA{Leave-one-out (LOO) sensitivity analysis for $p$, $q$, and $\eta$. $\mathrm{MAE}_\mathrm{full\text{-}fit}$ denotes the MAE obtained using all data points, while $\mathrm{MAE}_\mathrm{LOO}$ is the mean absolute error for the excluded data points predicted from the corresponding LOO fits. The maximum LOO error and the corresponding input variable $x_i$ are also reported. Here, $x_i = N_\mathrm{SIA}/L^2$ for $p$ and $q$, and $x_i = P/R_\mathrm{C}-6$ for $\eta$.}}
\label{tab:fit_sensitivity}

\begin{tabular}{ccccc}
\toprule
\revA{Quantity} & \revA{$\mathrm{MAE}_\mathrm{full\text{-}fit}$} & \revA{$\mathrm{MAE}_\mathrm{LOO}$} & \revA{$\max_i e_i^\mathrm{LOO}$} & \revA{$x_i$ at max. $e_i^\mathrm{LOO}$} \\
\midrule
\revA{$p$}    & \revA{0.252} & \revA{0.358} & \revA{0.945} & \revA{0.018} \\
\revA{$q$}    & \revA{0.128} & \revA{0.166} & \revA{0.482} & \revA{0.045} \\
\revA{$\eta$} & \revA{0.014} & \revA{0.035} & \revA{0.129} & \revA{0.161} \\
\bottomrule
\end{tabular}

\end{table}
}

\FloatBarrier


\revA{\section*{IX. Derivation of analytical expressions for the temperature dependence of $F$, $S$, and $\langle E_\mathrm{f} \rangle$}}

\revA{From the temperature dependence of $S(T)$ obtained from the statistical mechanics analysis (Fig.~\hl{10}~(c-1) in the main text), we observe that, for each $N_\mathrm{SIA}$, $S(T)$ exhibits an approximately linear dependence on $T$ in the high temperature range. In addition, its zero-temperature limit is determined by the degeneracy of the ground-state configurations, $S(0) = k_\mathrm{B} \ln g (E_\mathrm{f,ground})$. Motivated by these two features, we adopt the following empirical form:
    \begin{equation}
    S(T) = B_1T + B_2 \exp \left( - \frac{T}{B_3} \right) \text{,}
    \label{eq:s26}
    \end{equation}
where $B_1$, $B_2$, and $B_3$ are fitting parameters. The first term describes the approximately linear high-temperature behaviour, whereas the second term accounts for the zero-temperature configurational entropy and decays exponentially with increasing temperature, with $B_3$ characterising the associated temperature scale. Accordingly,
    \begin{equation}
    S(0) = B_2 = k_\mathrm{B} \ln g(E_\mathrm{f,ground}) \text{.}
    \label{eq:s27}
    \end{equation}
For a system with a finite excitation gap above the ground state, $S(T)$ is expected to approach its zero-temperature limit with a vanishing slope \cite{McQuarrie2000},
    \begin{equation}
    \frac{\partial S}{\partial T} \to 0 \qquad \text{as} \qquad T \to 0 \text{.}
    \end{equation}
Applying this condition to Eq. (\ref{eq:s26}) gives
    \begin{equation}
    B_1 - \frac{B_2}{B_3} = 0 \text{,}
    \end{equation}
and hence
    \begin{equation}
    B_1 = \frac{k_\mathrm{B} \ln g (E_\mathrm{f,ground})}{B_3} \text{.}
    \end{equation}
Therefore, Eq. (\ref{eq:s26}) can be expressed in terms of a single independent parameter $B_3$ as 
    \begin{equation}
    S(T) = k_\mathrm{B} \ln g (E_\mathrm{f,ground}) \left[ \exp \left( - \frac{T}{B_3} \right) + \frac{T}{B_3} \right] \text{.}
    \label{eq:s31}
    \end{equation}
We further assume that $B_3$ scales linearly with $\eta$, such that $B_3 = \tau \eta$, where $\tau$ is a parameter. The parameter $\eta$ characterises the deviation of the ground state geometry from the ideal hexagon (see the discussion in the main text), whereas $\tau$ represents a factor independent of $\eta$ that determines the characteristic temperature scale $B_3$. Substituting $B_3 = \tau \eta$ into Eq. (\ref{eq:s31}) then gives
    \begin{equation}
    S(T) = k_\mathrm{B} \ln g (E_\mathrm{f,ground}) \left[ \exp \left( - \frac{T}{\tau \eta} \right) + \frac{T}{\tau \eta} \right] \text{.}
    \label{eq:s32}
    \end{equation}
This expression is identical to Eq. (\hl{22}) in the main text.}

\revA{Using the thermodynamic relation $S = -\partial F / \partial T$, the analytical expression for $F(T)$ is obtained by integrating $S(T)$ with respect to $T$:
    \begin{equation}
    F(T) = E_\mathrm{f,ground} -\int_0^T S(T^{\prime}) \, dT^{\prime} \text{.}
    \label{eq:s33}
    \end{equation}
Substituting Eq. (\ref{eq:s32}) into Eq. (\ref{eq:s33}) yields
    \begin{equation}
    F (T) = k_\mathrm{B} \ln{g(E_\mathrm{f,ground})} \left[ \tau \eta \exp \left( -\frac{T}{\tau \eta} \right) - \frac{T^2}{2 \tau \eta} - \tau \eta \right] + E_\mathrm{f,ground}
    \end{equation}
which is identical to Eq. (\hl{20}) in the main text.}

\FloatBarrier

\vspace{2\baselineskip}


\section*{X. Molecular dynamics simulations for the glide motion of a dislocation loop}
We perform molecular dynamics simulations to investigate the effect of configurational variations in a dislocation loop on its glide motion. A cubic perfect-crystal cell with a side length of approximately 20.0 nm is oriented along the X $[111]$, Y $[\bar{2}11]$, and Z $[0\bar{1}1]$ directions. Periodic boundary conditions are applied in all directions. At the centre of the cell, we introduce a prismatic $1/2[111]$ dislocation loop containing $N_\mathrm{SIA}$ SIAs in its ground-state configuration, identified by Wang--Landau sampling as described in the main text. The system energy is then minimised using the conjugate gradient method. The molecular dynamics simulations are performed in the NPT ensemble using a Nose--Hoover thermostat and barostat. The temperature is increased to the target value at a rate of 5 K $\mathrm{ps^{-1}}$ and subsequently maintained constant, while the pressure is kept at zero throughout the simulations. The target temperature ranges from 70 to 800 K. For each condition, we track the trajectory of the loop centre of mass along the [111] direction for 40 ns. The trajectory is decomposed into 40 segments of 1 ns each, and the diffusion coefficients are averaged over all segments \cite{Zhao2017ActaMater}. The activation energy for glide motion ($E_{a,\mathrm{glide}}$) and the prefactor ($D_{0,\mathrm{glide}}$) are evaluated by applying the Arrhenius law to the temperature dependence of the averaged diffusion constant. The simulations are carried out using the large atomic/molecular massively parallel simulator (LAMMPS) \cite{Plimpton1995}.

    \begin{figure}[htbp]
    \centering
    \includegraphics[width=0.9\linewidth]{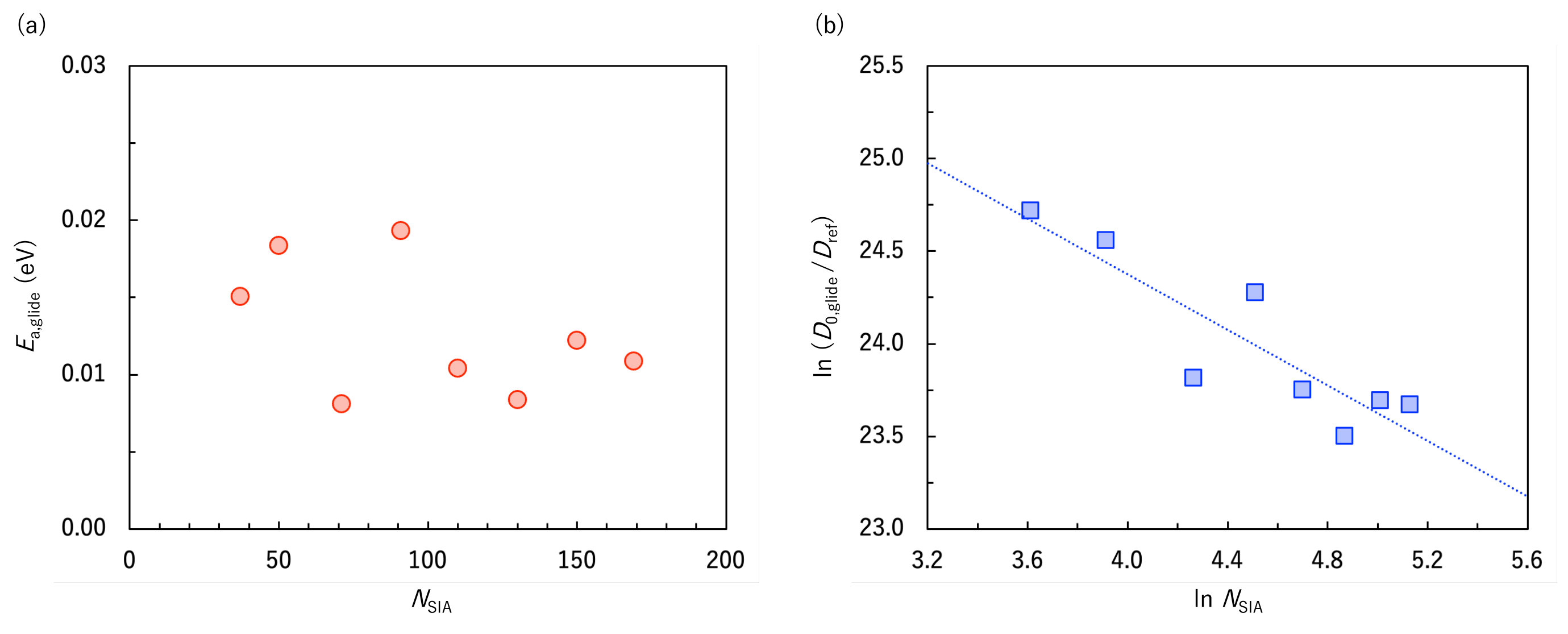}
    \caption{(a) Effective activation energy $E_{a,\mathrm{glide}}$ and prefactor $D_{0,\mathrm{glide}}$ for loop glide as functions of $N_\mathrm{SIA}$. $D_\mathrm{ref}$ denotes the reference diffusion constant used for normalisation, with $D_\mathrm{ref} =$ 1 $\mathrm{nm}^2$ $\mathrm{s}^{-1}$.}
    \label{fig:one-d_motion}
    \end{figure}

Fig. \ref{fig:one-d_motion} (a) shows $E_{a,\mathrm{glide}}$ as a function of $N_\mathrm{SIA}$, indicating no discernible trend. It should be noted that the $N_\mathrm{SIA}$ values considered here are the same as those used in Fig. \hl{9} (e) of the main text, including large configurational variations characterised by the parameter $\eta$. Fig. \ref{fig:one-d_motion} (b) shows $D_{0,\mathrm{glide}}$ as a function of $N_\mathrm{SIA}$, demonstrating a power-law dependence of $D_{0,\mathrm{glide}}$ on $N_\mathrm{SIA}$:
    \begin{equation}
    D_{0,\mathrm{glide}} = D_{0,\mathrm{glide}}^* N_\mathrm{SIA}^\beta \text{,} \label{eq:s21}
    \end{equation}
where $D_{0,\mathrm{glide}}^*$ and $\beta$ are fitting parameters, with values $D_{0,\mathrm{glide}}^* = 7.746 \times 10^{11}$ $\mathrm{nm}^2$ $\mathrm{s}^{-1}$ and $\beta = -0.750$, respectively. Previous molecular dynamics simulations have also reported a power-law dependence for loop glide in bcc Fe and face-centred cubic Cu \cite{Osetsky2000JNuclMater_276_65}. Based on these results, we conclude that loop configurational variation has little influence on glide motion; specifically, $E_{a,\mathrm{glide}}$ does not depend on the loop configuration or $N_\mathrm{SIA}$, whereas $D_{0,\mathrm{glide}}$ is dominantly controlled by $N_\mathrm{SIA}$.

\section*{XI. Simulation setup for a nudged elastic band calculation}

    \begin{figure}[b]
    \centering
    \includegraphics[width=0.7\linewidth]{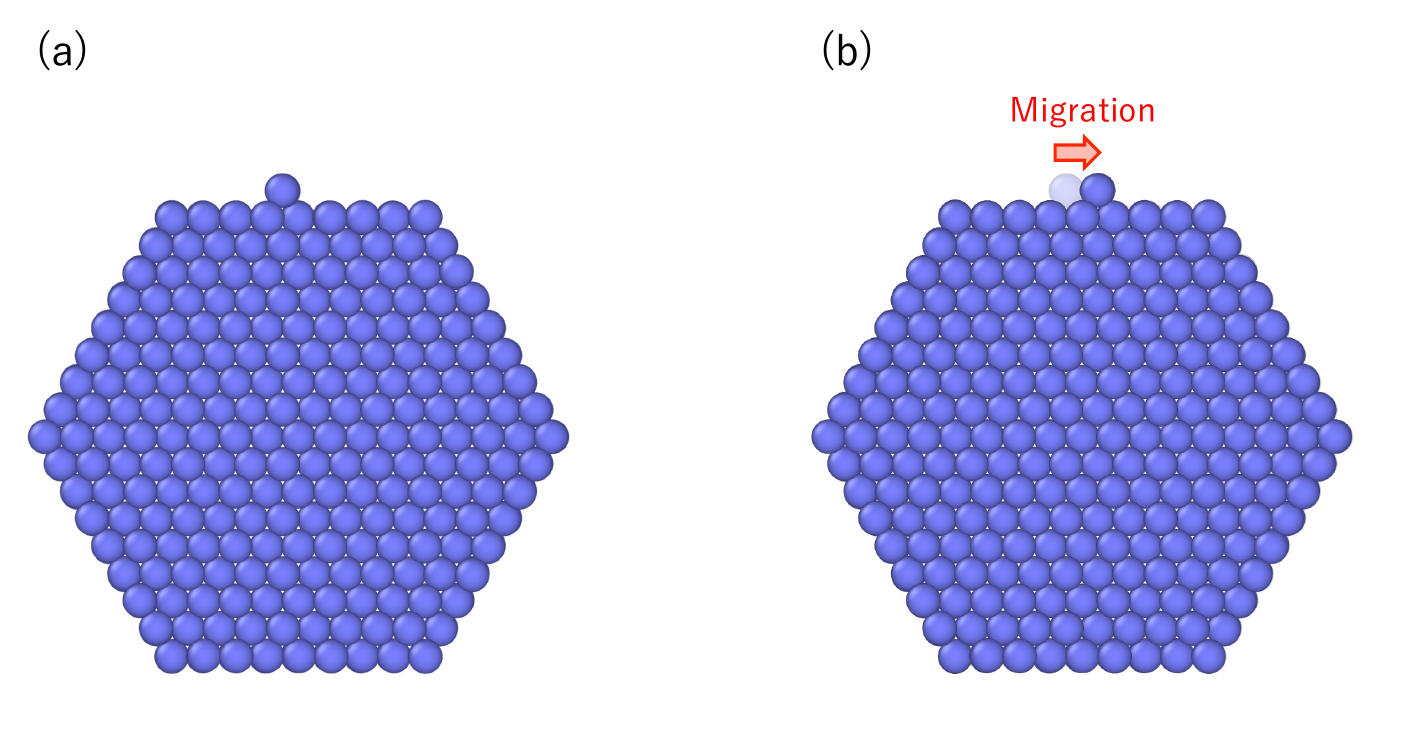}
    \caption{Configurations of a dislocation loop with an additional SIA located at the loop edge: (a) initial configuration before migration and (b) final configuration after migration. The arrow in (b) indicates the migration direction of the additional SIA along the loop edge.}
    \label{fig:NEB}
    \end{figure}
    
The value of $E_\mathrm{m}$ in Eq. (\hl{24}) in the main text corresponds to the energy barrier for pipe-diffusion along the loop edge in the absence of a change in $E_\mathrm{f}$ before and after the migration.
To obtain this value, we construct a perfectly hexagonal dislocation loop with $N_\mathrm{SIA} = 217$ in a $20 \times 20 \times 20$ $\mathrm{nm}^3$ simulation cell and place an additional SIA at the edge such that the migration does not alter $E_\mathrm{f}$, as illustrated in Fig. \ref{fig:NEB}.
The migration barrier for this process is calculated using the climbing-image nudged elastic band method.
The transition pathway is discretized into 8 images, and atomic relaxations are performed using the fast inertial relaxation engine (FIRE) algorithm, implemented in LAMMPS \cite{Plimpton1995}.
The minimisation is terminated when the maximum two-norm of the force vector on each replica falls below 0.02 eV $\mathrm{eV} \, \mathrm{A}^{-1}$.
This calculation yields a migration barrier of $E_\mathrm{m} = 2.359$ eV.

\bibliography{references}